\documentclass[12pt, draftclsnofoot, onecolumn]{IEEEtran}
\normalsize
\usepackage{pifont,bm,multicol,amsfonts,amsmath,color,amssymb,graphicx,amsthm, epsfig,cite,psfrag,subfigure,algorithm,enumerate,stfloats,algorithm,algorithmic,epstopdf,balance}

\newtheorem{lemma}{\underline{Lemma}}

\newtheorem{remark}{\underline{Remark}}

\begin{document}
\title{A New Path Division Multiple Access for the Massive MIMO-OTFS Networks} 	

\author{ Muye Li, Shun Zhang, \emph{Member, IEEE}, Feifei Gao, \emph{Fellow, IEEE}, Pingzhi Fan, \emph{Fellow, IEEE}, Octavia~A.~Dobre,~\emph {Fellow, IEEE}

\thanks{M. Li, S. Zhang are with the State Key Laboratory of Integrated Services Networks, Xidian University, Xi¡¯an 710071, P. R. China (e-mail: myli$\_$96@stu.xidian.edu.cn; zhangshunsdu@xidian.edu.cn).}

    \thanks{F. Gao is with Department of Automation, Tsinghua University, State Key Lab of Intelligent Technologies and Systems, Tsinghua University, State Key for Information Science and Technology (TNList) Beijing 100084, P. R. China (e-mail: feifeigao@ieee.org).}

    \thanks{P. Fan is with the Southwest Jiaotong University, Chengdu 611756, P. R. China (e-mail: pzfan@swjtu.edu.cn).}

    \thanks{O. A. Dobre is with Faculty of Engineering and Applied Science, Memorial University of Newfoundland, St. John's NLAIC-5S7, Canada (e-mail: odobre@mun.ca).}

}
\maketitle
  \begin{abstract}
This paper focuses on a new path division multiple access (PDMA) for both uplink (UL) and downlink (DL) massive multiple-input multiple-output network over a high mobility scenario, where the orthogonal time frequency space (OTFS) is adopted.
First, the 3D UL channel model and the received signal model in the angle-delay-Doppler domain are studied.
Secondly, the 3D-Newtonized orthogonal matching pursuit algorithm is utilized for the extraction of the UL channel parameters, including channel gains, directions of arrival, delays, and Doppler frequencies, over the antenna-time-frequency domain.
Thirdly, we carefully analyze
energy dispersion and power leakage of the 3D angle-delay-Doppler channels. Then, along UL,
we design a path scheduling algorithm to properly assign angle-domain resources at user sides and to assure that
the observation regions for different users do not overlap over the 3D cubic area, i.e.,  angle-delay-Doppler domain.
After scheduling, different users can
map their respective data to the scheduled delay-Doppler domain grids,
and simultaneously send the data to base station (BS) without inter-user interference in
the same OTFS block. Correspondingly, the signals at desired grids within the 3D resource space of BS are separately collected to implement the 3D channel estimation and maximal ratio combining-based data detection over the angle-delay-Doppler domain.
Then, we construct a low complexity beamforming scheme over the angle-delay-Domain domain to achieve inter-user interference free DL communication.
Simulation results are provided to demonstrate the validity of our proposed unified UL/DL PDMA scheme.

\end{abstract}

\maketitle
\thispagestyle{empty}


\section{Introduction}

Massive multiple-input multiple-output (MIMO) is a promising technology for the next generation communication networks.
Once the base station (BS) is equipped with a large number of antennas to serve multi-users, the significant gains in terms of both energy and capacity efficiency can be achieved\cite{massive_MIMO,Massive_in_5G_1,4}.
Similar to the other communication systems, the channel estimation is vital for the massive MIMO networks.
With accurate channel state information, we can analyze the system achievable rate \cite{achie_rate,2}, quantize the network interference \cite{interference_quanti},
and derive the energy efficiency of the system \cite{energy_efficiency}.
Then, we can optimize the signal processing units of massive MIMO, such as
beamforming \cite{beamforming}, user scheduling \cite{user_grouping}, data detection \cite{data_detection}.
Thus, the effective channel acquisition over massive MIMO has become a hot topic.
From various measurement campaigns about massive MIMO channels, we can find that the scattering effect of the environment is limited in one narrow angle spread region \cite{angle_domain}.
Then, researchers have fully exploited this characteristic and proposed efficient channel recovery schemes~\cite{channel_recovery2, channel_recovery3,channel_recovery_GaoZhen}.

However, in higher practice, the users may move, and the channels would vary in time.
As is well known, the higher the mobility speed is, the less the channel coherence time is.
Then, it would be more challenging to acquire a large number of unknown channels within a limited channel coherence time.
Correspondingly, this problem has attracted much attention, and some interesting and important results have been presented in the literatures as follows.
With respect to the time-selective massive MIMO channel, Xie \emph{et al.} firstly proposed the spatial basis expansion model (SBEM) for the representation of the uplink (UL)/ downlink (DL) channels in \cite{gao_E}.
In \cite{Ma_J_SBL_Time_varing}, Ma \emph{et al.} proposed the auto-regressive (AR) model to measure the time-varying channel. They developed an expectation maximization (EM) based sparse Bayesian learning (SBL) framework to learn the temporal and spatial channel parameters.
In \cite{MYLi_SBL_Time_varing}, Li \emph{et al.} extended the work in \cite{Ma_J_SBL_Time_varing}, and proposed an optimal Bayesian Kalman filter-based channel tracking method with only partial prior knowledge of the DL channel parameters.
Zhao \emph{et al.} designed a channel tracking method for massive MIMO systems under both time-varying and spatial-varying conditions in \cite{channel_recovery1}.
Under the double-selective massive MIMO channel fading scenario,
Qin \emph{et al.} proposed an effective time-varying channel estimation scheme for the massive MIMO-orthogonal frequency division multiplex (OFDM) system, where the complex exponential BEM (CE-BEM) was utilized in \cite{Q_Qin}.
Zhang \emph{et al.} presented a novel recovery algorithm based on sparsity adaptive matching pursuit for compressed sensing-based sparse channel estimation in OFDM systems in \cite{Octavia_OFDM}.
In \cite{You_OFDM}, You \emph{et al.} proposed adjustable phase shift pilots for time-varying channel acquisition in a massive MIMO-OFDM system.
In \cite{Timevarying_Guvensen}, Guvensen \emph{et al.} gave a pre-beamformer design method for the spatially correlated time-varying wideband MIMO channels under the assumption that the channel is a stationary Gauss-Markov random process.
Hu \emph{et al.} considered the angle-domain Doppler shifts compensation for high-mobility massive MIMO communication in \cite{high_mobility4}.

Generally, the above works can be split into two categories: AR model- and BEM-based schemes.
Within the former framework, the massive MIMO channels are usually assumed
to be block fading, and the time-varying characteristics within the block would be lost.
Although the latter can characterize the massive MIMO channel dynamic features within the given block,
the parameters for BEM are sensitive to the user mobility conditions \cite{BEM1}. Moreover, there exists model approximation errors in both models, especially under high speed conditions \cite{BEM2}.
The reasons for the above phenomena is that the two methods are not from the time-varying physical scattering model and can not capture intrinsic factors, which effectively quantize and describe the time-varying channels.

Recently, Hadani \emph{et al.}
designed a novel two-dimensional modulation technique called orthogonal time frequency space
(OTFS) modulation, and gave us a new perspective on time-varying channels \cite{OTFS_original}.
In the traditional OFDM schemes, the effective data are mapped and processed
over the time-frequency domain. Thus, we usually insert pilots
and estimate the time-varying channels within the time-frequency domain.
However, OTFS provides a new two-dimensional signal space, i.e., the delay-Doppler domain.
Correspondingly, the data mapping and demapping are implemented over the delay-Doppler domain.
Interestingly, the time-varying channels could be equivalently described by
several constant variables over the delay-Doppler domain.
In \cite{IC_OTFS}, Raviteja \emph{et al.} developed Hadani's work by considering
the time-varying physical scattering model, and derived the explicit input-output model over
the delay-Doppler domain. Importantly, Raviteja proved that the equivalent channels over
the delay-Doppler domain are determined by the parameters of each scattering path, such as delay and Doppler frequency.
Raviteja \emph{et al.} developed embedded pilot-aided OTFS channel estimation schemes in \cite{OTFS5}.
In \cite{OTFS_MA}, Khammammetti \emph{et al.} proposed a multiple access method in the UL of an OTFS system, where the users are allocated delay-Doppler resource blocks which are spaced at equal intervals in the domain.

\begin{figure*}[t]
  \centering
  \includegraphics[width=140mm]{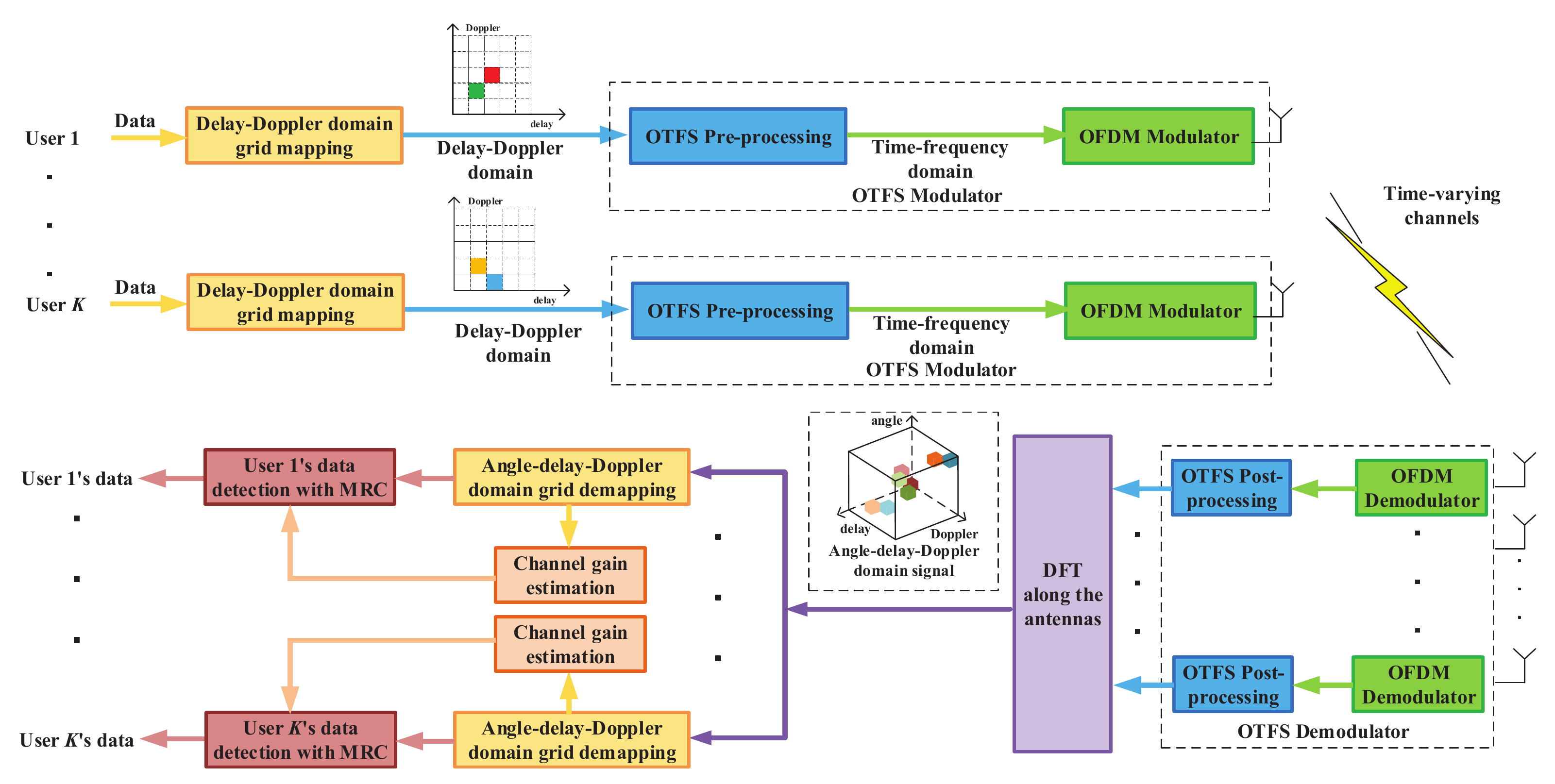}\\
  \caption{The block diagram of the proposed PDMA scheme for the massive MIMO-OTFS system. DFT: discrete Fourier transform.}\label{fig_framework}
\end{figure*}

However, the important characteristic of OTFS is that many signal blocks will be received with different double circular shift structures over the delay-Doppler domain; accordingly, a low complexity data detection method is not easy to design \cite{OTFS_dd1, OTFS_dd2}.
In this paper, we fully utilize the super scattering path resolution of the massive MIMO and examine the massive MIMO-OTFS over the 3D signal space, i.e.,  angle-delay-Doppler domain.
In fact, Shen \emph{et al.} studied the OTFS modulation for massive MIMO systems, but only considered channel estimation \cite{proof}.
Here, within the time division duplex (TDD) model, we propose a unified path division multiple access (PDMA) scheme for both UL and DL over massive MIMO-OTFS.
The high mobility UL channel model and the OTFS signal model in the angle-delay-Doppler domain are developed.
Then, we resort to the 3D Newtonized orthogonal matching pursuit (NOMP) algorithm to recover the UL channel parameters, including channel gains, direction of arrival (DOAs), delays, and Doppler frequencies, over the antenna-time-frequency domain.
Correspondingly, both energy dispersion and power leakage phenomena of the 3D angle-delay-Doppler channels are carefully analyzed.
With respect to UL, we design a path scheduling algorithm to properly assign angle-domain resources at user sides and to assure that
the observation regions for different users do not overlap over the 3D cubic area, i.e.,  angle-delay-Doppler domain.
After scheduling, different users can map their respective data to the scheduled delay-Doppler domain grids, and simultaneously send the data to the BS without inter-user interference in the same OTFS block.
Then, the signals at desired grids within the 3D resource space of the BS are separately collected to implement the 3D channel estimation and maximal ratio combining (MRC)-based data detection over the angle-delay-Doppler domain.
Furthermore, we apply the idea of UL's path scheduling algorithm for the multi-user service along DL and design a low complexity beamforming strategy for the massive MIMO-OTFS DL over the angle-delay-Doppler domain.
The block diagram for the UL part of our proposed PDMA scheme is illustrated in~\figurename{ \ref{fig_framework}}.

The rest of this paper is organized as follows.
Section II describes the 3D channel model.
Section III introduces the UL channel parameter extraction with 3D-NOMP.
The main ideas of the UL/DL PDMA scheme are presented in Section IV.
Simulation results are shown in Section V, and conclusions are drawn in Section VI.

Notations: Denote lowercase (uppercase) boldface as vector (matrix).
$(\cdot )^H $, $(\cdot )^T $, and $(\cdot )^{*} $ represent the Hermitian, transpose, and conjugate, respectively.
$\mathbf I_N $ is an $N \times N $ identity matrix.
$\mathbb E \{\cdot \} $ is the expectation operator.
Denote $\text{tr} \{\cdot \}$ and $|\cdot | $ as the trace and the determinant of a matrix, respectively.
$[\mathbf A]_{i,j} $ and $\mathbf A_{\mathcal Q,:}$ (or $\mathbf A_{:, \mathcal Q} $) represent the $(i,j) $-th entry of $\mathbf A $ and the submatrix of $\mathbf A $ which contains the rows (or columns) with the index set $\mathcal Q $, respectively.
$\mathbf x_{\mathcal Q} $ is the subvector of $\mathbf x $ built by the index set $\mathcal Q $.
$\mathbf v \sim \mathcal{CN} (\mathbf 0, \mathbf I_N)$ means that $\mathbf v $ follows the complex Gaussian distribution with zero mean and covariance $\mathbf I_N $.
$\lfloor p \rfloor $ denotes the largest integer less than or equal to $p $.
The real component of $x $ is expressed as $\Re (x)$.
$\text{diag} (\mathbf x)$ is a diagonal matrix whose diagonal elements are formed with the elements of $\mathbf x $.

\section{System Model}
\subsection{High Mobility Massive MIMO Channel Model}
We consider a single-cell mmWave massive MIMO system in high-mobility scenarios.
The BS serves $K$ users, which are randomly distributed in the area.
The BS is equipped with a uniform linear array (ULA), which contains $N_r$ antenna elements and $N_r \gg K$.
Each user has a single antenna element.
There are scatterers in the space and the user channels are composed of multiple propagation paths.
The wireless signal can reach the BS along the line of sight path or can be reflected by multiple scatterers, which means that the channel links between the BS and the users is subject to frequency-selective fading.
Due to the high mobility of the users, the channels vary fast and experience time-selective fading.
It is assumed that there are $P$ scattering paths for the channel between a specific user and the BS.
Each scattering path corresponds to one DOA, one Doppler frequency shift, and one time delay.
%

Denote $\theta_{k,p} (n)$ as the DOA for the $p$-th path of the $k$-th user at time block $n$.
The corresponding spatial steering vector can be defined as:
\begin{align}
\mathbf{a}(\!\theta_{k,p} (\!n))\!=\! [1, e^{\jmath 2\pi \frac{d \!\sin(\!\theta_{k,p} \! (n) ) }{\lambda}}\!,\! \ldots\!,\!e^{\jmath 2\pi (N_r\!-\!1\!) \frac{d\!\sin(\theta_{k,p} \! (n))}{\lambda}}]^T,
\end{align}
where  $\lambda$  is the carrier wavelength for the UL, and the antenna spacing $d$ is set as half the wavelength $\lambda$.
Hence, from the geometric channel model, the time-varying UL channel for the $k$-th user at time block $n$ can be denoted by
\begin{align}
\mathbf h_{k,l}(n)=\sum_{p=1}^P {h}_{k,p}e^{\jmath2\pi\nu_{k,p}nT_s} \delta(lT_s-\tau_{k,p})
\mathbf{a}(\theta_{k,p} (n)),\label{eq:h_kln}
\end{align}
where $h_{k,p}$, $\tau_{k,p}$ and $\nu_{k,p}$ are the channel gain, delay, and Doppler shift for the $p$-th path of the user $k$, respectively,
$\delta(\cdot)$ denotes the Dirac function, and $T_s$ is the system sampling period.
Fractional delays are not considered since the resolution of  $T_s$ is sufficient to capture the detailed channel information along the delay dimension over the typical wide-band systems.
Hence, we assume that
 $\tau_{k,p}=n_{\tau,p} T_s$, where $n_{\tau,p}$ is an integer number.
As $\theta_{k,p}(n)$ remains constant during a long time interval \cite{Ma_J_SBL_Time_varing,gao_E,MYLi_SBL_Time_varing}, the time index $n$ of the angle can be omitted.
Furthermore, $h_{k,p} \sim \mathcal{CN}(0, \lambda_{k,p}) $, and $\boldsymbol \nu_k = [\nu_{k,1}, \nu_{k,2}, \ldots, \nu_{k,P}]^T$, $\boldsymbol \tau_k = [\tau_{k,1}, \tau_{k,2}, \ldots, \tau_{k,P}]^T$, and $\boldsymbol \theta_k = [\theta_{k,1}, \theta_{k,2}, \ldots, \theta_{k,P}]^T$ for further use.

\subsection{Massive MIMO-OTFS Scheme}
For the $k$-th user, we rearrange a data sequence of length $L_D N_D$ into a two-dimensional OTFS data block $\mathbf{X}_{k} \in \mathcal{C}^{L_D \times N_D}$, where $L_D$ and $N_D$ are the dimensions of the delay domain and the Doppler domain, respectively.

First, we apply the inverse symmetric finite Fourier transform (ISFFT) for the  pre-processing block, and obtain the data block $\tilde{\mathbf{X}}_{k}$  in the time-frequency domain as
$\tilde{\mathbf{X}}_{k}=\mathbf{F}_{L_D}\mathbf{X}_{k}\mathbf{F}_{N_D}^H$.
Notice that $\mathbf{F}_{L_D}\in\mathcal{C}^{L_D\times L_D}$ and $\mathbf{F}_{N_D}\in\mathcal{C}^{N_D\times N_D}$ are
normalized DFT matrices with entries $[\mathbf F_{n}]_{p,q}=\frac{1}{\sqrt{n}} e^{-\jmath\frac{2\pi pq}{n}}$, $p,q=0,1,2,\ldots,n-1$, $n=L_D, N_D$.
Then, we take the $L_D$-point inverse DFT (IDFT) on each column of $\tilde{\mathbf{X}}_{k}$ and obtain the transmitting signal block $\mathbf {S}_{k} =[\mathbf s_{k,0},\mathbf s_{k,2},\ldots,\mathbf s_{k,N_D-1}]=\mathbf{F}_{L_D}^H \tilde{\mathbf{X}}_{k}$,
where $\mathbf s_{k,j}$ denotes an OFDM symbol.

By adding the cyclic prefix (CP) for each OFDM symbol,
we can obtain the one-dimensional transmitting signal $\mathbf{s}_{k} \in \mathcal {C}^{(L_{cp} + L_D) N_D\times 1}$ over time domain,
where $L_{cp}$ is the length of CP.
Then, $\mathbf{s}_{k}$ will occupy the bandwidth $L_D \bigtriangleup f$ with duration $N_D T$, where $\bigtriangleup f$ and $T=(L_{cp} + L_D)T_s$ are the subcarrier spacing and the OFDM symbol period, respectively.

At the BS, we sequentially implement the symmetric operation with that at users, such as the rearranging, removing CP, $L_D$-point DFT and
the SFFT operation of size $L_D\times N_D$ in the post-processing block (see Fig. \ref{fig_framework}).
Correspondingly, at the $n_r$-th antenna of the BS, we can obtain
the two-dimensional data block $\mathbf{Y}_{n_r}\in\mathcal{C}^{L_D \times N_D}$ in the delay-Doppler domain.
%
From \cite{OTFS_model}, the $(i, j+ N_D/2)$-th entry of $\mathbf Y_{n_r}$ can be denoted as
\begin{align}
[\mathbf Y_{n_r}]_{i, j+N_D/2} =& \sum_{k=1}^{K} \sum_{i^{\prime}=0}^{L_D - 1} \sum_{j^{\prime} = -N_D / 2}^{N_D/2-1} [\mathbf X_k]_{i^{\prime}, j^{\prime}+N_D/2}
\sum_{n=1}^{N_D}
e^{-j 2 \pi(n-1) \frac{j-j^{\prime}}{N_D}}
\notag \\
&\times \frac{1}{N_D} \left[\mathbf{h}_{k, (i-i^{\prime})_{L_D}}\left((n-1)\left(L_{D}+L_{c p}\right)+i+1\right)\right]_{n_r}
+  w_{i,j,n_r}
,
\label{eq:real Y^DD1}
\end{align}
where $(i-i')_{L_D}$ is the remainder by $L_D$ after division of $i-i'$, $i=0,1,\ldots,L_D-1$ and $j=-N_D/2,\ldots,0,\ldots,N_D/2-1$.
Moreover, $w_{i,j}$ is complex Gaussian noise with zero mean and variance $\sigma_n^2$, and is independent from element to element.

\begin{lemma}\label{lemma1}

The $(i, j+ N_D/2)$-th entry of $\mathbf Y_{n_r}$ can be re-expressed as
\begin{align}
[\mathbf Y_{n_r}]_{i, j+N_D/2} =& \sum_{k=1}^{K} \sum_{i^{\prime}=0}^{L_D - 1} \sum_{j^{\prime} = -N_D / 2}^{N_D/2-1} [\mathbf X_k]_{i^{\prime}, j^{\prime}+N_D/2} \notag \\
&\times\left(\tilde{h}_{k,(i-i^{\prime})_{L_D}, \langle j-j'\rangle, n_r}
+ \tilde{g}_{k,(i-i^{\prime})_{L_D}, \langle j-j'\rangle, n_r}^{i} \right)
+ w_{i,j,n_r},
\label{eq:real Y^DD2}
\end{align}
where $\langle j-j'\rangle = (j-j^{\prime}+ N_D/2)_{N_D}-N_D/2$ and
\begin{small}\begin{align}
&\tilde{h}_{k,i,j,n_r}= \frac{1}{N_D}\!\sum_{p=1}^P \!h_{k,p}e^{\jmath2\pi\nu_{k,p}T_s}
\Upsilon_{N_D}(\nu_{k,p} N_D T\!-\!j) \delta(iT_s-\tau_{k,p})
e^{\jmath2 \pi n_r \frac{d\sin \theta_{k,p}}{\lambda}},\label{eq:main H^DDS} \\
&\tilde{g}_{k,i,j,n_r}^{\ell} = \frac{1}{N_D}\!\sum_{p=1}^P \! 2\jmath e^{\jmath \pi \nu_{k,p} \ell T_s } \sin(\pi \nu_{k,p} \ell T_s)
h_{k,p} e^{\jmath2\pi\nu_{k,p}T_s}
\Upsilon_{N_D}(\nu_{k,p} N_D T\!-\!j) \delta(iT_s-\tau_{k,p})
e^{\jmath2 \pi n_r \frac{d\sin \theta_{k,p}}{\lambda}} \label{eq:auxiliary H^DDS}
\end{align}\end{small}are separately the main and the secondary channels over the space-delay-Doppler domain.
Moreover, $\ell$ is the coordinate of the received grid along the delay dimension at the BS,
and $\Upsilon_N(x) \triangleq \sum\limits_{n=1}^N e^{\jmath 2\pi \frac{x}{N}(n-1)} = \frac{\sin(\pi x)}{\sin(\pi \frac{x}{N})} e^{\jmath \pi \frac{x(N-1)}{N}}$.
\begin{proof}
Refer to Appendix A.
\end{proof}
\end{lemma}


To further take into account the channel sparsity caused by the massive antennas, we can utilize the spatial DFT operation along the antenna index $n_r$.
By applying the normalized DFT along the antenna index $n_r$,
we can derive the angle-delay-Doppler domain channel $\bar{h}_{k,(i,j,q)}$ and $\bar{g}_{k,(i,j,q)}^{\ell}$ as
\begin{small}\begin{align}
\bar{h}_{k,(i,j,q)}
=& \frac{1}{N_D\sqrt N_r} \sum_{p=1}^P
h_{k,p} e^{\jmath2 \pi\nu_{k,p}T_s}
\mathcal A(\boldsymbol \chi_{k,p}, i,j,q), \label{eq:fiveB H^DDA} \\
\bar{g}_{k,(i,j,q)}^{\ell}
=& \frac{1}{N_D\sqrt N_r} \sum_{p=1}^P
2\jmath e^{\jmath \pi \nu_{k,p} \ell T_s } \sin(\pi \nu_{k,p} \ell T_s)
h_{k,p} e^{\jmath2 \pi\nu_{k,p}T_s}
\mathcal A(\boldsymbol \chi_{k,p}, i,j,q), \label{eq:fiveB g^DDA}
\end{align}\end{small}where $\mathcal A(\boldsymbol \chi_{k,p}, i,j,q) = \Upsilon_{N_D}(\nu_{k,p} N_D T\!-\!j)
\delta (iT_s-\tau_{k,p})
\Upsilon_{N_r} (N_r \frac{d \sin \theta_{k,p}}{\lambda}-q)$,
$\boldsymbol \chi_{k,p} = [\nu_{k,p}, \tau_{k,p}, \theta_{k,p}] $,
and $q=-\frac{N_r}{2},\ldots,0,\ldots,\frac{N_r}{2}-1$.

\begin{figure*}[t]
  \centering
  \includegraphics[width=60mm]{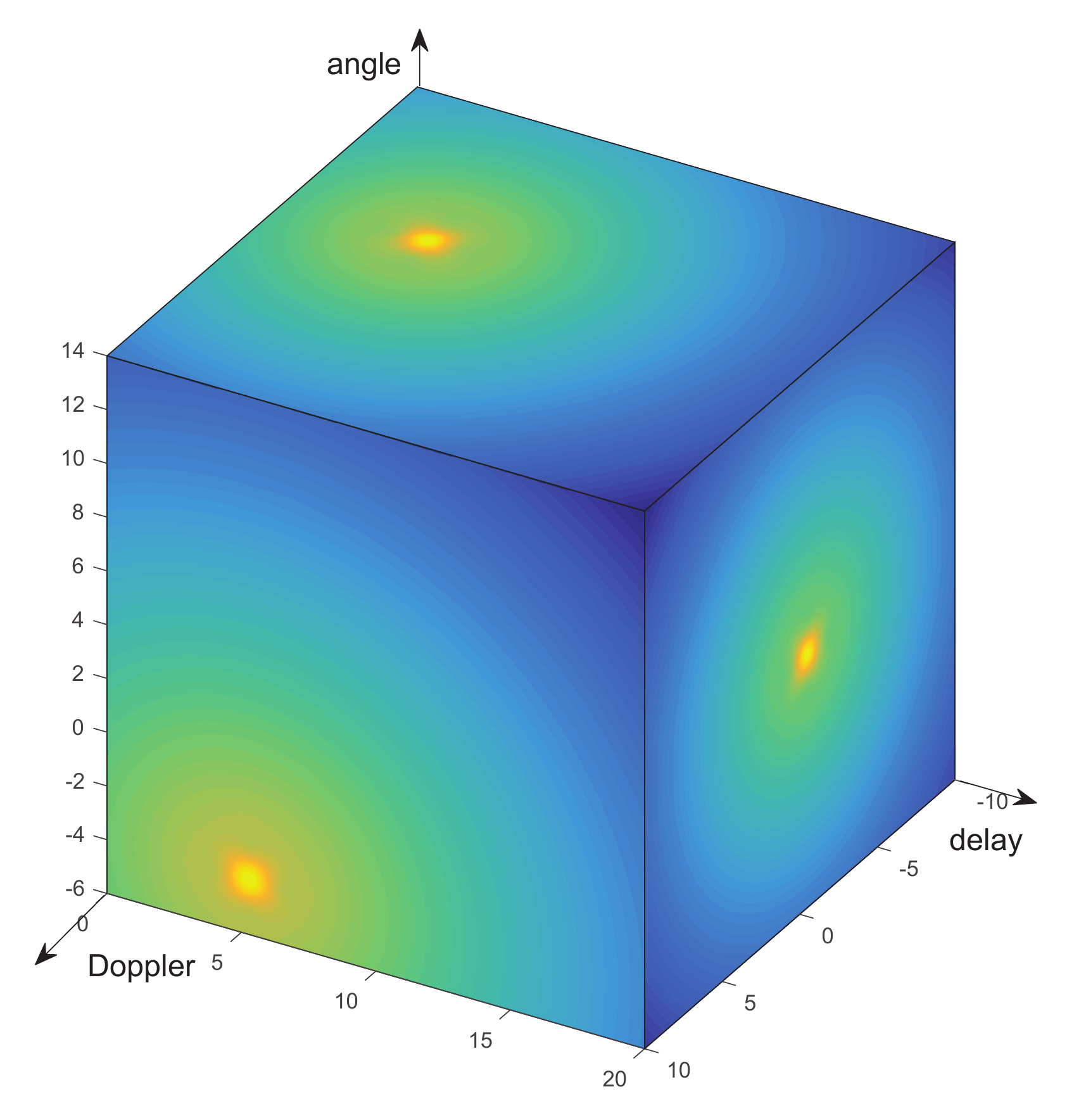}\\
  \caption{An illustration of the angle-delay-Doppler domain channel.}
  \label{fig_3D_Channel}
\end{figure*}

From \eqref{eq:fiveB H^DDA} and \eqref{eq:fiveB g^DDA}, it can be seen that the main channels $\bar{h}_{k,(i,j,q)}$ are only related to the channel parameters $\{\tau_{k,p}, \nu_{k,p}, \theta_{k,p}, h_{k,p}\}_{p=1}^P$,
while the secondary channels $\bar{g}_{k,(i,j,q)}^{\ell}$ are related to the coordinate of the received grid along the delay domain.
Besides, $\bar{h}_{k,(i,j,q)}$ and $\bar{g}_{k,(i,j,q)}^{\ell}$ have dominant elements only if $i\approx\tau_{k,p} L_D \Delta f$, $j\approx\nu_{k,p}N_D T$, and $q \approx N_r\frac{d\sin\theta_{k,p}}{\lambda}$, and each dominant element corresponds to a specific parameter set $\{\tau_{k,p}, \nu_{k,p}, \theta_{k,p}, h_{k,p}\}$.
Therefore, there are only $P$ dominant values within the $N_r L_D N_D$ elements, which means that $\bar{h}_{k,(i,j,q)}$ and $\bar{g}_{k,(i,j,q)}^{\ell}$ are sparse over the angle-delay-Doppler domain.
An illustration of the 3D domain channel is shown in Fig. \ref{fig_3D_Channel}.

By applying DFT to for the received signal $\mathbf y_{i,j+N_D/2}$, we can derive
\begin{align}
\bar{y}_{i,j+N_D/2, q + N_r/2}
=& [\mathbf F_{N_r} \mathbf {y}_{i,j+N_D/2}]_{q} \notag \\
=& \sum_{k=1}^{K}\sum_{i'=0}^{L_D-1} \!\!\sum_{j'=-N_D/2}^{N_D/2-1} \! [\mathbf X_{k}]_{i',j'+N_D/2} \notag \\
&\times (\bar{h}_{k,((i \!-\! i')_{L_D}, \langle j-j'\rangle, q)}
+ \bar{g}_{k,((i \!-\! i')_{L_D},\langle j-j'\rangle, q)}^{i} ) \!\!+\! \bar w_{i,j \!+ N_D/2, q+N_r/2},
\label{eq:fiveA Y^DD}
\end{align}
where the equivalent noise $\bar w_{i,j, q} = [\mathbf F_{N_r} \mathbf w_{i,j}]_{q}$ is still a white Gaussian noise with zero mean and variance $\sigma_n^2$.

\begin{remark}
Both $\bar{h}_{k,(i,j,q)}$ and $\bar{g}_{k,(i,j,q)}^{\ell}$ are formed by $P$ path components,
which can be seen from (\ref{eq:fiveB H^DDA}) and (\ref{eq:fiveB g^DDA}).
Given a specific path, the difference between the component in $\bar{h}_{k,(i,j,q)}$ and that in
$\bar{g}_{k,(i,j,q)}^{\ell}$ lies in the coefficient $2\jmath e^{\jmath \pi \nu_{k,p} \ell T_s } \sin(\pi \nu_{k,p} \ell T_s)$, which is related with the position of the grid in the delay domain at the receiver and the Doppler frequency of the path.
Nonetheless, if the parameter sets $\{\tau_{k,p}, \nu_{k,p}, \theta_{k,p}, h_{k,p} \}_{p=1}^P$ can be accurately estimated, we can simultaneously determine the positions of the $P$ dominant values for both $\bar{h}_{k,(i,j,q)}$ and $\bar{g}_{k,(i,j,q)}^{\ell}$, and reconstruct their explicit values.
It is worth noting that the dominant values for the main and the secondary channels are at the same grids in the angle-delay-Doppler domain.
\end{remark}
%

\section{Uplink Channel Parameter Capture with 3D-NOMP}

From the received signal model \eqref{eq:fiveA Y^DD}, it is obvious that the goal of the channel estimation is to obtain the  parameter sets $\{\tau_{k,p}, \nu_{k,p}, \theta_{k,p}, h_{k,p}\}_{p=1}^P$ for each user.
In this section, we construct the UL training model and adopt the 3D NOMP algorithm to estimate the parameter sets.

\subsection{UL Training Model}
The UL channel of the $k$-th user can be directly expressed with the $P$ parameter sets as \eqref{eq:h_kln}.
Different users implement their channel parameter estimation within the time division duplex (TDD) model.
Without loss of generality, we assume that each user utilizes the same training sequence $\mathbf t_{cp} = \left[ t_{N_{t} - L_{c p}}, t_{N_{t} - L_{c p}+1}, \ldots, t_{N_{t} - 1}, \mathbf{t}^{T} \right]^{T} \in \mathcal{C}^{\left(L_{c p + N_{t}}\right) \times 1} $,
where $L_{cp}$ is the CP length, and $N_t$ is the number of effective points.
Furthermore, $\mathbf {t}=\left[t_{0}, t_{1}, \ldots, t_{N_{t}-1}\right]^{T}$, and $\|\mathbf{t}\|^{2} = P_t $ is the power of the training sequence.
The first user starts its training at time $n_1 T_s$, and the length of training for each user is $(L_{cp} + N_t) T_s $.
Then, we can derive that the $k$-th user will start its training at $\left(n_{1}+\left(L_{c p}+N_{t}\right)(k-1)\right) T_{s}$.
At the BS, we cast away the samples corresponding to the CPs and collect the receiving signals at $\left(n_{1}+\left(L_{c p}+N_{t}\right)(k-1)+L_{c p}+n\right) T_{s}$ for the $k$-th user into the $N_r \times 1$ vector $\mathbf y_{k,n}$ as
\begin{align}\label{receive_est}
\mathbf{y}_{k, n} =\sum_{p=1}^{P} h_{k, p} e^{\jmath 2 \pi \nu_{k, p} \left( n_{1} + \left( L_{cp} + N_{t} \right)(k-1) + L_{cp} + n\right) T_{s}}
t_{\left(n-\tau_{k, p} / T_{s}\right)_{N_{t}}} \mathbf{a}\left(\theta_{k, p} \right) + \mathbf{w}_{k, n},
\end{align}
where $n = 0,1, \ldots, N_t-1$.
Define the number of experienced slots $f_{k,n} =  n_{1} + L_{cp} + N_{t} (k-1) + L_{cp} + n$, the Doppler phase bias vectors $\mathbf v(\nu_{k,p}) = [e^{\jmath 2 \pi \nu_{k, p} f_{k,0} T_{s}}, \ldots, e^{\jmath 2 \pi \nu_{k, p} f_{k,N_t-1} T_{s}}]^T $ and the training vectors $\mathbf t_d(\tau_{k,p}) = [t_{\left(0-\tau_{k, p} / T_{s}\right)_{N_{t}}}, \ldots, t_{\left((N_t-1)-\tau_{k, p} / T_{s}\right)_{N_{t}}}]^T$.
Then, the received signal can be expressed as
\begin{align}\label{receive_est_vec}
\mathbf{y}_{k} =& \sum_{p=1}^{P} h_{k, p} (\mathbf v(\nu_{k,p}) \odot
\mathbf t_d(\tau_{k,p})) \otimes \mathbf{a}\left(\theta_{k, p} \right) + \mathbf{w}_{k}
= \sum_{p=1}^{P} h_{k, p} \mathbf p(\theta_{k, p}, \tau_{k,p}, \nu_{k,p}) + \mathbf{w}_{k},
\end{align}
where $\mathbf{w}_{k} = [\mathbf{w}_{k, 0}^T, \ldots, \mathbf{w}_{k, N_t-1}^T]^T$ and $\mathbf p(\theta_{k, p}, \tau_{k,p}, \nu_{k,p}) = (\mathbf v(\nu_{k,p}) \odot
\mathbf t_d(\tau_{k,p})) \otimes \mathbf{a}\left(\theta_{k, p} \right)$.

After the construction of the received signal model \eqref{receive_est_vec},
the next task is to estimate the parameter sets $\{\theta_{k,p}, \tau_{k,p}, \nu_{k,p}, h_{k,p}\}_{p=1}^P$.
In the next subsection, we resort to the 3D-NOMP algorithm for the estimation process.

\subsection{3D-NOMP Algorithm}
For our model, we extend the original 2D-NOMP algorithm \cite{ori_NOMP} to 3D.
Firstly, the 3D coarse searching is implemented.
The next step is to precisely search near the result of the first step.
Then, the single refinement for the parameters at the current iteration and the cyclic refinement for the results of the past iterations are executed.
Finally, the gains updating is carried out through the least square (LS) algorithm.
The detailed steps in the $i$-th iteration of the 3D-NOMP algorithm are given in the following.

\subsubsection{3D-coarse searching}
The ranges of actual $\theta_{k,i}$, $\tau_{k,i}$, and $\nu_{k,i}$ are $[-\frac{\pi}{2}, \frac{\pi}{2}]$,
$[0, \tau_{max}] $,
and $[-\nu_{max}, \nu_{max}] $,
respectively, where $\nu_{max} = \frac{v_{max}}{\lambda}$ and $\tau_{max} = N_{\tau}\leq (\frac{L_D}{2} -1) T_s$, with $v_{max}$ as the maximum permitted velocity in the system.
To reduce the searching complexity, we construct the low-dimension 3D under-sampled map $\mathbf \Omega_l $ for the domains of angle, delay and Doppler frequency as
\begin{align}\label{coarse_map}
\mathbf \Omega_l=&\left\{\left(k_{1}, k_{2}, k_{3}  \right), k_{1} = -\frac{N_r / \eta_{\theta}}{2}, \ldots, 0, \ldots, \frac{N_r / \eta_{\theta}}{2} - 1; \right. \notag \\
&\left. k_2 = 0, \ldots, N_t-1;
k_3 = -\frac{N_D / \eta_{\nu}}{2}, \ldots, 0, \ldots, \frac{N_D / \eta_{\nu}}{2} -1 \right\},
\end{align}
where $\eta_{\theta}$ and $\eta_{\nu}$ are the under-sampling rates for the domain of angle and Doppler, respectively.
For one grid $(k_1, k_2, k_3)$ in $\mathbf \Omega_l$, the corresponding parameters $({\theta}, {\tau}, {\nu})$ can be derived by $(\frac{\pi}{2} \frac{k_1}{N_r/ 2\eta_{\theta}}, k_2 T_s, \frac{k_3 \nu_{max}}{N_D / 2\eta_{\nu}} )$.
The coarsely estimated $\bar{\theta}_{k,p}$, $\bar{\tau}_{k,p}$, $\bar{\nu}_{k,p}$  are selected as
\begin{align}\label{searching1}
\left(\bar{\theta}_{k,i}, \bar{\tau}_{k,i}, \bar{\nu}_{k,i}\right) =
\arg \max _{(k_1, k_2, k_3) \in \mathbf \Omega_l} \frac{\left| \mathbf p^H (\theta, \tau, \nu) \mathbf{y}_{k,r}^{i-1}\right|^{2}}{\|\mathbf p(\theta, \tau, \nu)\|^{2}},
\end{align}
where $\mathbf y_{k,r}^{i-1}$ is the residual at the end of the $(i-1)$-th iteration, calculated as:
\begin{align}\label{residual}
\mathbf{y}_{k,r}^{i-1} = \mathbf{y} - \sum_{p=1}^{i-1} \hat{h}_{k,p}
\mathbf p(\hat{\theta}_{k,p}, \hat{\tau}_{k,p}, \hat{\nu}_{k,p}),
\end{align}
and $\{\hat\theta_{k,p}, \hat\tau_{k,p}, \hat\nu_{k,p}, \hat h_{k,p}\}_{p=1}^{i-1}$ are the estimated parameters in the previous $i-1$ iterations.

\subsubsection{3D-precisely searching for $\nu_{k,p}$ and $\theta_{k,p} $}
After the simultaneously coarse searching for the three parameters,
we do a similar step, i.e., 3D-precisely searching, near the results of  the coarse searching step.
And the details are omitted here for length limitation.
Then, the parameters $\tilde{\theta}_{k,i}, \tilde{\tau}_{k,i}, \tilde{\nu}_{k,i}$ can be obtained through a similar operation in (\ref{searching1}).
Afterwards, the coarse estimate of the channel gain $\tilde{h}_{k,i}$ can be expressed as:
\begin{align}\label{updating_h}
\tilde{h}_{k,i} = \frac{\mathbf p^H (\tilde{\theta}_{k,i}, \tilde{\tau}_{k,i}, \tilde{\nu}_{k,i}) \mathbf{y}_{k,r}^{i-1}}{\left\| \mathbf p(\tilde{\theta}_{k,i}, \tilde{\tau}_{k,i}, \tilde{\nu}_{k,i})\right\|^{2}}.
\end{align}

\subsubsection{Single refinement}
During the last step, the precise estimation of delay $\hat\tau_{k,i} = \tilde{\tau}_{k,i}$ is performed.
Here, we will resort to the extended Newton method and refine $\tilde\theta_{k,i}$, $\tilde\nu_{k,i}$, and $\tilde h_{k,i}$.
$R_s$ iterations are be executed in this step.
The goal of the refinement step is to minimize the power of the new residual $\left\|\mathbf{y}_{k,r} - h \mathbf{p}(\theta, \tau, \nu)\right\|^{2}$ in the $i$-th iteration.
Hence, the objective is to maximize the following function:
\begin{align}\label{objective}
{J}\left(h, \theta, \nu\right) =2 \Re \left\{\mathbf{y}_{k,r}^H h \mathbf{p}(\theta, \hat\tau_{k,i}, \nu)\right\} - \left| h\right|^{2}\|\mathbf{p}(\theta, \hat\tau_{k,i}, \nu)\|^{2}.
\end{align}
Then, the refined estimates of $\theta_{k,i}$ and $\nu_{k,i}$ can be represented as
\begin{align}\label{refine1}
\left[\begin{matrix}{\tilde{\theta}'_{k,i}} \\
{\tilde{\nu}'_{k,i}}\end{matrix}\right]
=\left[\begin{matrix}{\tilde{\theta}_{k,i}} \\
{\tilde{\nu}_{k,i}}\end{matrix}\right]
- \mathbf{J}^{\prime\prime} \left(\tilde{h}_{k,i}, \tilde{\theta}_{k,i}, \tilde{\nu}_{k,i}\right)^{-1}
\mathbf{J}^{\prime} \left(\tilde{h}_{k,i}, \tilde{\theta}_{k,i}, \tilde{\nu}_{k,i}\right),
\end{align}
where
\begin{align}\label{first_order_all}
\mathbf{J}^{\prime} \left({\tilde h_{k,i}}, \tilde{\theta}_{k,i}, \tilde{\nu}_{k,i} \right)
= \left[ \begin{matrix}
{\frac{\partial J}{\partial \tilde\theta_{k,i}}} \\
{\frac{\partial J}{\partial \tilde\nu_{k,i}}}
\end{matrix}\right]
\!=\! \left[ \begin{matrix}
{2 \Re\left\{\tilde h_{k,i} \left(\mathbf{y}_{k,r}^{i-1} - \tilde h_{k,i} \mathbf{p} \right)^{H} \frac{\partial \mathbf{p}}{\partial \tilde{\theta}_{k,i}} \right\}} \\
{2 \Re\left\{\tilde h_{k,i} \left(\mathbf{y}_{k,r}^{i-1} - \tilde h_{k,i} \mathbf{p} \right)^{H} \frac{\partial \mathbf{p}}{\partial \tilde{\nu}_{k,i}} \right\}}
\end{matrix}\right]
\end{align}
is the first-order partial derivative vector, with
$\frac{\partial \mathbf{p}}{\partial \tilde{\theta}_{k,i}} = (\mathbf v(\tilde\nu_{k,i}) \odot
\mathbf t_d(\hat\tau_{k,i})) \otimes \mathbf{a}^{\prime} (\tilde\theta_{k, i} )$,
$\frac{\partial \mathbf{p}}{\partial \tilde{\nu}_{k,i}} = (\mathbf v^{\prime} (\tilde\nu_{k,i}) \odot
\mathbf t_d(\hat\tau_{k,i})) \otimes \mathbf{a}(\tilde\theta_{k, i} )$,
and
the $n_r$-th element of $\mathbf{a}^{\prime}(\tilde\theta_{k,i})$ and the $n_t$-th element of $\mathbf v^{\prime} (\tilde \nu_{k,i})$ are $[\mathbf{a}^{\prime}(\tilde\theta_{k,i})]_{n_r} = {\jmath 2\pi n_r \frac{d \cos(\tilde\theta_{k,i})}{\lambda}} e^{\jmath 2\pi n_r \frac{d \sin(\tilde\theta_{k,i})}{\lambda}}$,
$[\mathbf v^{\prime} (\tilde \nu_{k,i})]_{n_t} = \jmath 2 \pi f_{k,n_t-1} T_{s} e^{\jmath 2 \pi \tilde\nu_{k, i} f_{k,n_t-1} T_{s}}$, respectively, with $n_r=0,1,\ldots, N_r-1$ and $n_t=0,1,\ldots, N_t-1$.

In addition, the second-order partial derivative matrix in \eqref{refine1} can be derived as
\begin{align}\label{second_order_all}
\mathbf{J}^{\prime \prime} \left({\tilde h_{k,i}}, \tilde{\theta}_{k,i}, \tilde{\nu}_{k,i} \right)
= \left[\begin{matrix}
{\frac{\partial^{2} J}{\partial \tilde{\theta}_{k,i}^2 }} &
{\frac{\partial^{2} J}{\partial \tilde{\theta}_{k,i} \partial \tilde{\nu}_{k,i} }} \\
{\frac{\partial^{2} J}{\partial \tilde{\nu}_{k,i} \partial \tilde{\theta}_{k,i} }} &
 {\frac{\partial^{2} J}{\partial \tilde{\nu}_{k,i}^2 }}
\end{matrix}\right],
\end{align}
where
\begin{align}
\frac{\partial^{2} J}{\partial \tilde{\theta}_{k,i}^2 }&
\!=\!2 \Re \left\{\tilde h_{k,i} \left(\mathbf {y}_{k,r}^{i-1} - \tilde h_{k,i} \mathbf{p}\right)^{H} \!\!(\mathbf v(\tilde\nu_{k,i}) \odot \mathbf t_d(\hat\tau_{k,i})) \otimes \mathbf{a}^{\prime \prime} (\tilde\theta_{k, i} )
- \left|\tilde h_{k,i} \right|^{2} \left\|\frac{\partial \mathbf{p}}{\partial \tilde{\theta}_{k,i}} \right\|^{2} \right\}, \\
\frac{\partial^{2} J}{\partial \tilde{\theta}_{k,i} \partial \tilde{\nu}_{k,i} }&
\!=\!2 \!\Re\! \left\{\tilde h_{k,i} \!\!\left(\mathbf {y}_{k,r}^{i-1} \!-\! \tilde h_{k,i} \mathbf{p}\right)^{H} \!\!(\mathbf v^{\prime} (\tilde\nu_{k,i}) \!\odot \!\mathbf t_d(\hat\tau_{k,i})) \otimes \mathbf{a}^{\prime} (\tilde\theta_{k, i} )
\!-\! \left|\tilde h_{k,i} \right|^{2} (\frac{\partial \mathbf{p}}{\partial \tilde{\nu}_{k,i}})^{H} \!\frac{\partial \mathbf{p}}{\partial \tilde{\theta}_{k,i}}\right\}, \\
\frac{\partial^{2} J}{\partial \tilde{\nu}_{k,i} \partial \tilde{\theta}_{k,i} }&
\!=\! 2 \!\Re\! \left\{\tilde h_{k,i} \!\!\left(\mathbf {y}_{k,r}^{i-1} \!-\! \tilde h_{k,i} \mathbf{p}\right)^{H} \!\!(\mathbf v^{\prime} (\tilde\nu_{k,i}) \odot \mathbf t_d(\hat\tau_{k,i}))\! \otimes \!\mathbf{a}^{\prime} (\tilde\theta_{k, i} )
\!-\! \left|\tilde h_{k,i} \right|^{2} (\frac{\partial \mathbf{p}}{\partial \tilde{\theta}_{k,i}})^{H} \!\frac{\partial \mathbf{p}}{\partial \tilde{\nu}_{k,i}}\right\}, \\
\frac{\partial^{2} J}{\partial \tilde{\nu}_{k,i}^2 }&
\!=\! 2 \Re \left\{\tilde h_{k,i} \left(\mathbf {y}_{k,r}^{i-1} - \tilde h_{k,i} \mathbf{p}\right)^{H}\!\! (\mathbf v^{\prime \prime} (\tilde\nu_{k,i}) \odot \mathbf t_d(\hat\tau_{k,i})) \otimes \mathbf{a} (\tilde\theta_{k, i} )
- \left|\tilde h_{k,i} \right|^{2} \left\|\frac{\partial \mathbf{p}}{\partial \tilde{\nu}_{k,i}} \right\|^{2}\right\},
\end{align}
and the $n_t$-th element of $\mathbf v^{\prime\prime} (\tilde \nu_{k,i})$ is
$[\mathbf v^{\prime\prime} (\tilde \nu_{k,i})]_{n_t} = - (2 \pi f_{k,n_t} T_{s})^2 e^{\jmath 2 \pi \tilde\nu_{k, i} f_{k,n_t} T_{s}}$.
Moreover, the $n_r$-th element of $\mathbf{a}^{\prime\prime}(\tilde\theta_{k,i})$
is $[\mathbf{a}^{\prime\prime}(\tilde\theta_{k,i})]_{n_r}=
({-\jmath 2\pi n_r \frac{d \sin(\tilde\theta_{k,i})}{\lambda} - (2\pi n_r \frac{d \cos(\tilde\theta_{k,i})}{\lambda})^2 }) e^{\jmath 2\pi n_r \frac{d \sin(\tilde\theta_{k,i}) }{\lambda}}$.


By carrying out \eqref{refine1}, $\tilde\theta_{k,i}$ and $\tilde\nu_{k,i}$ are refined.
The gain $\tilde h_{k,i}$ is then updated according to \eqref{updating_h}.

\subsubsection{Cyclic refinement}
After the single refinement step for the parameters of the current iteration, to further perfect the estimates $\{\tilde h_{k,p}, \tilde\theta_{k,p}, \tilde\nu_{k,p}\}_{p=1}^{i-1}$ of the previous iterations, $R_c$ iterations of cyclically refinement are taken into consideration.
Similar to the single refinement step, the extended Newton method is also utilized;
details are omitted here due to space limitation.
Together with the acquired estimates of the delay in the searching step, accurate estimates of $\tilde h_{k,p}$, $\hat\theta_{k,p}$, $\hat\tau_{k,p}$, $\hat\nu_{k,p}$, $p = 1, 2, \ldots, i$ are obtained.

\subsubsection{Gain updating}
With the estimated $\hat\theta_{k,p}$, $\hat\tau_{k,p}$, $\hat\nu_{k,p}$ in the previous $i$ iterations, we can update the channel gain $h_{k,p}$, $p = 1, 2, \ldots, i$ with the LS estimator as
\begin{align}
[\hat h_{k,1}, \hat h_{k,2}, \ldots, \hat h_{k,i}]^T = \mathbf P^{\dagger} \mathbf y_k,
\end{align}
where $\mathbf P = [\mathbf p(\hat{\theta}_{k,1}, \hat{\tau}_{k,1}, \hat{\nu}_{k,1}), \ldots, \mathbf p(\hat{\theta}_{k,i}, \hat{\tau}_{k,i}, \hat{\nu}_{k,i})]$.

\subsubsection{Stopping criterion}
As the NOMP runs, the power of the residual $\mathbf y_{k,r}$ decreases after each iteration.
If the parameters are accurate enough, all the actual paths of the channel are extracted,
and the power of the residual is reduced to the power of the noise after the final iteration, i.e., $\|\mathbf y_{k,r}\|^2 \approx \|\mathbf w_k\|^2 $.
Hence, the 3D-NOMP algorithm terminates when
\begin{align}
\left|\mathbf{p}(\theta, \tau, \nu)^{H} \mathbf {y}_{k,r}^i \right|^2 < \epsilon
\end{align}
for all possible $(\theta, \tau, \nu)$.
%
The stopping criterion threshold $\epsilon$ is chosen from the false alarm rate
$P_{fa} = P\{\|\mathbf w_k\|_{\infty}^2 > \epsilon\} = 1 - (1- \exp (-\epsilon/\sigma_n^2))^{N_r N_t}$ as \cite{Jin_NOMP}
\begin{align}
\epsilon = -\sigma_n^{2} \ln (1-(1-P_{\mathrm{fa}})^{1 / N_r N_t}).
\end{align}

\section{PDMA over the Angle-Delay-Doppler Domain}

After the acquisition of the UL channel parameter sets $\{\tau_{k,p}, \nu_{k,p}, \theta_{k,p}, h_{k,p}\}_{p=1}^P$, the next step is the data transmission and detection.
As the distances between the users and BS are much farther than the moving distance of the user within a number of OTFS blocks, the angles and the distance changes of the scattering paths are negligible within a period of time.
For example, within a OTFS system, we set  the  number of subcarriers as $L_D = 512$,  the number of OFDM blocks as $N_D = 128$, the  length of CP as $L_{cp} = 32$, and the sampling rate as $1/T_s = 20$ MHz. Then, if the user moves with the speed $360$ km/h and the distance between BS and the user is $500$ m,
the maximal angle change within $5$ OTFS blocks is $0.2^\circ$, which means that the parameter sets $\{\tau_{k,p}, \nu_{k,p}, \theta_{k,p}\}_{p=1}^P$ can be treated as constant variables in $5$ OTFS blocks.
Then,  knowledge about $\{\tau_{k,p}, \nu_{k,p}, \theta_{k,p}\}_{p=1}^P$ from the previous section can be utilized for this section.
However, as the OTFS block contains several OFDM symbols, the channel gain $h_{k,p}$ may change from one OTFS block to another one.
Nonetheless, there is no need to directly track the original channel gain $h_{k,p}$.
Instead, we estimate the equivalent channel gain $\bar{h}_{k,(i,j,q)}$ and $\bar{g}_{k,(i,j,q)}^{\ell}$ over the angle-delay-Doppler domain.
We first propose an UL path scheduling algorithm, and then simultaneously estimate the channel gain and detect the transmitted data.
Finally, a low-complexity DL beamforming strategy is presented with the similar idea of UL scheduling.

\subsection{Analysis of 3D Channel over the Angle-Delay-Doppler Domain}

From \eqref{eq:fiveB H^DDA} and \eqref{eq:fiveB g^DDA},
it can be checked that $\bar{h}_{k,(i,j,q)}$ and $\bar{g}_{k,(i,j,q)}^{\ell}$ are dominant at the index set
$\mathcal Q_k=\left\{ (i_{k,p},j_{k,p},q_{k,p})\right\}_{p=1}^P$, where
\begin{align}
&q_{k,p} = \lfloor N_r \frac{d \sin \theta_{k,p}}{\lambda} \rfloor, \kern 10pt
i_{k,p} = \lfloor \tau_{k,p} L_D \triangle f \rfloor, \kern 10pt
j_{k,p} = \lfloor \nu_{k,p}N_DT \rfloor. \label{eq:signature}
\end{align}

Obviously, $(i_{k,p},j_{k,p},q_{k,p})$ corresponds to the $p$-th physical scattering path of the $k$-th user and can be treated as the delay-Doppler-angle signature of this path.
Moreover, the $p$-th path contains almost the entire channel power at the dominant grid $(i_{k,p},j_{k,p},q_{k,p})$. Hence, at the grid $(i_{k,p},j_{k,p},q_{k,p})$, we have the following approximation
\begin{align}
\bar{h}_{k,(i_{k,p},j_{k,p},q_{k,p})} \approx&
\frac{1}{\sqrt N_r}
h_{k,p} e^{\jmath2 \pi\nu_{k,p}T_s}
\mathcal A(\boldsymbol \chi_{k,p}, i_{k,p},j_{k,p},q_{k,p}),\label{eq:approx H^DDA} \\
\bar{g}_{k,(i_{k,p},j_{k,p},q_{k,p})}^{\ell} \approx&
\frac{1}{\sqrt N_r}
2\jmath e^{\jmath \pi \nu_{k,p} \ell T_s } \sin(\pi \nu_{k,p} \ell T_s)
h_{k,p} e^{\jmath2 \pi\nu_{k,p}T_s}
\mathcal A(\boldsymbol \chi_{k,p}, i_{k,p},j_{k,p},q_{k,p}),\label{eq:approx g^DDA}
\end{align}
where $\ell$ has been defined in the previous section as the received grid index over the delay domain.
Correspondingly, the following relation can be written
\begin{align}\label{relation_h_g}
\bar{g}_{k,(i_{k,p},j_{k,p},q_{k,p})}^{\ell} \approx&
2\jmath e^{\jmath \pi \nu_{k,p} \ell T_s }
\sin(\pi \nu_{k,p} \ell T_s)
\bar{h}_{k,(i_{k,p},j_{k,p},q_{k,p})}.
\end{align}

With (\ref{eq:fiveA Y^DD}) and (\ref{eq:approx H^DDA})--(\ref{relation_h_g}), we have the following observations.
Within a specific OTFS block, the main channels $\bar{h}_{k,(i,j,q)}$ are constant over the
angle-delay-Doppler domain, and each scattering path corresponds to one main channel gain. On the other hand, the secondary channels $\bar{g}_{k,(i,j,q)}^{\ell}$ are varying at different observation grids. Nonetheless, with the parameter sets $\{\tau_{k,p}, \nu_{k,p}, \theta_{k,p}\}_{p=1}^P$, $\bar{g}_{k,(i_{k,p},j_{k,p},q_{k,p})}^{\ell}$ can be well constructed
from the main channels $\bar{h}_{k,(i_{k,p},j_{k,p},q_{k,p})}$ at different observation grids.
In order to clearly illustrate the impact of the secondary channels, we consider a typical massive MIMO-OTFS example,
where the carrier frequency is $30$ GHz, the sampling rate is $1/T_s = 20 $ MHz, the number of the grids along the delay domain is $L_D =256$,
and the number of grids along the Doppler frequency is $N_D=64$. We consider two mobility scenarios: the low speed
$36$ km/h and the high speed $360$ km/h, whose maximum Doppler frequencies are
$1 $ kHz and $10 $ kHz, respectively. Then, we consider the grids with the delay domain index $\ell=50$,
and set $\nu_{k,p}$ as the maximum Doppler shift.
With respect to the coefficient $\sin(\pi \nu_{k,p} \ell T_s)$ in (\ref{relation_h_g}), it can be
checked that $\sin(\pi \nu_{k,p} \ell T_s) = 8\times10^{-3}$ for the low speed scenario and
$\sin(\pi \nu_{k,p} \ell T_s) = 7.85 \times 10^{-2}$ under the high speed case. Obviously, the
power of the secondary channels can not be ignored for the mobility scenario.

In practice, $L_D$ and $N_r$ are large but not infinite, and unavoidable power leakage occurs, which is due to the characteristic of the function $\text{Sa}_N(x)$ \cite{gao_E}.
The neighboring grids of the dominant grid in the 3D space may be contaminated.
Let us analyze the leaked power of the grid $(i_{k,p}, j_{k,p}+d^{\nu}, q_{k,p}+d^{\theta})$
from the dominant channel grid $(i_{k,p},j_{k,p},q_{k,p})$, where $d^{\nu}$ and $d^{\theta}$ are the distance along Doppler and angle domains, respectively.
Then, the observed leaked channel power from the main and secondary channels can be
written as $|\bar{h}_{k,(i_{k,p},\langle j_{k,p}+d^{\nu} \rangle, q_{k,p}+d^{\theta})}+ \bar{g}_{k,(i_{k,p}, \langle j_{k,p}+d^{\nu} \rangle,q_{k,p}+d^{\theta})}^{\ell}|^2$.
From \eqref{eq:fiveB H^DDA} and \eqref{eq:fiveB g^DDA}, we define the leakage ratio $\eta_{k,p}(d^{\nu}, d^{\theta})$  as
\begin{align}
\eta_{k,p}(d^{\nu}, d^{\theta})
=& \frac{ \mathbb E_{h}\{|{\bar{h}}_{k, (i_{k,p}, \langle j_{k,p}+d^{\nu} \rangle, q_{k,p}+d^{\theta})} + {\bar{g}}_{k, (i_{k,p}, \langle j_{k,p}+d^{\nu} \rangle, q_{k,p}+d^{\theta})}^{\ell}|^2\} }
{
\mathbb E_{h} \{|{\bar{h}}_{k, (i_{k,p}, j_{k,p}, q_{k,p})} + {\bar{g}}_{k, (i_{k,p}, j_{k,p}, q_{k,p})}^{\ell}|^2\} } \notag \\
= &\frac{\sum\limits_{p'=1}^{P} \left|
\mathcal A(\boldsymbol \chi_{k,p'}, i_{k,p}, \langle j_{k,p}+d^{\nu} \rangle,q_{k,p}+d^{\theta})\right|^2 \lambda_{k,p'}
}
{\sum\limits_{p'=1}^{P} \left|
\mathcal A(\boldsymbol \chi_{k,p'}, i_{k,p}, j_{k,p}, q_{k,p})\right|^2 \lambda_{k,p'}
}.
\end{align}
Since almost the entire power on the dominant grid $(i_{k,p}, j_{k,p}, q_{k,p})$ comes from the $p$-th scattering path, we can further approximate $\eta_{k,p}(d^{\nu}, d^{\theta})$ as
\begin{align}\label{interference_ratio}
\eta_{k,p}(d^{\nu}, d^{\theta})
\approx &\sum_{\substack{p'=1\\ \tau_{k,p'} = \tau_{k,p}}}^{P}
\left|\frac{
\frac{\sin(\pi(\nu_{k,p'} N_D T\!-\!(j_{k,p}+d^{\nu})))}{\sin(\pi\frac{\nu_{k,p'} N_D T\!-\!(j_{k,p}+d^{\nu})}{N_D})}}
{\frac{\sin(\pi(\nu_{k,p} N_D T\!-\!j_{k,p}))}{\sin(\pi\frac{\nu_{k,p} N_D T\!-\!j_{k,p}}{N_D})}}\right|^2
\left|\frac{
\frac{\sin(\pi(N_r \frac{d \sin \theta_{k,p'}}{\lambda}-(q_{k,p}+d^{\theta})))}
{\sin(\pi\frac{(N_r \frac{d \sin \theta_{k,p'}}{\lambda}-(q_{k,p}+d^{\theta}))}{N_r})}}
{\frac{\sin(\pi(N_r \frac{d \sin \theta_{k,p}}{\lambda}-q_{k,p}))}
{\sin(\pi\frac{(N_r \frac{d \sin \theta_{k,p}}{\lambda}-q_{k,p})}{N_r})}}\right|^2
\frac{\lambda_{k,p'}}{\lambda_{k,p}}.
\end{align}

\begin{figure}[!t]
  \centering
  \includegraphics[width=80mm]{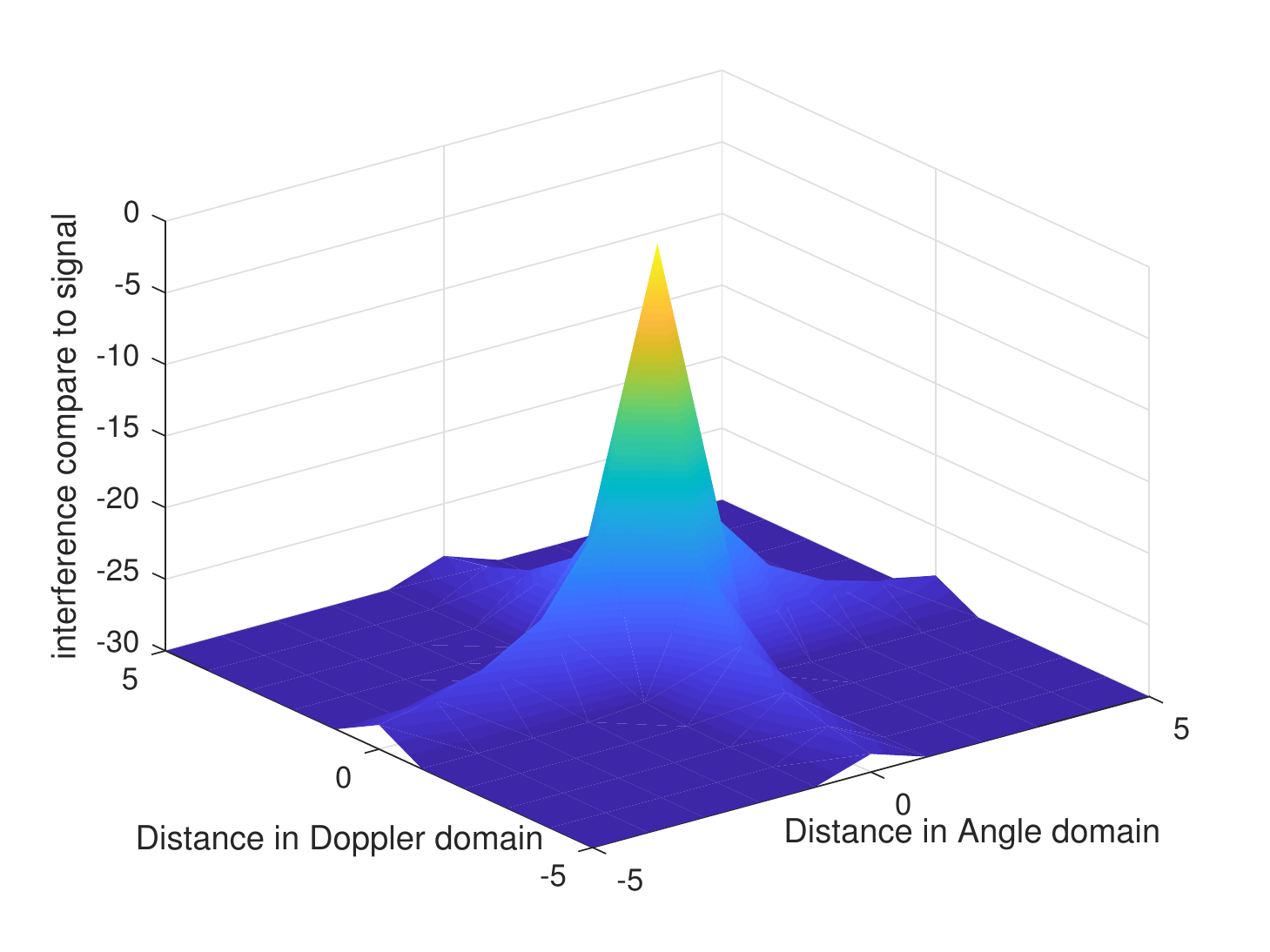}\\
  \caption{The interference vs. the distance of the grids on the axis of angle and Doppler at SNR = 30dB.} \label{fig:interference_measure}
\end{figure}

A simple example for the channel path with $\theta_{k,p} = 34^{\circ}$, user velocity $360$ km/h, $N_r = 128$, $L_D = 128$, and $N_D = 128$ is given in \figurename{ \ref{fig:interference_measure}}, where the signal-to-noise ratio (SNR) is set as 30 dB.
The parameter $\nu_{k,p}$ is set as the maximum Doppler shift.
From \figurename{ \ref{fig:interference_measure}}, it can be observed that $\eta_{k,p} (d^{\nu}, d^{\theta})$ is only -20 dB
when the geometric distance between the observed grid and the corresponding dominant grid is $d^{\nu\theta} = d^{\nu} + d^{\theta} = 1$.
Furthermore, if $d^{\nu\theta} >1 $, $\eta_{k,p}(d^{\nu}, d^{\theta})$ becomes small, and the power leakage almost has no influence.

\begin{figure*}[!t]
  \centering
  \includegraphics[width=110mm]{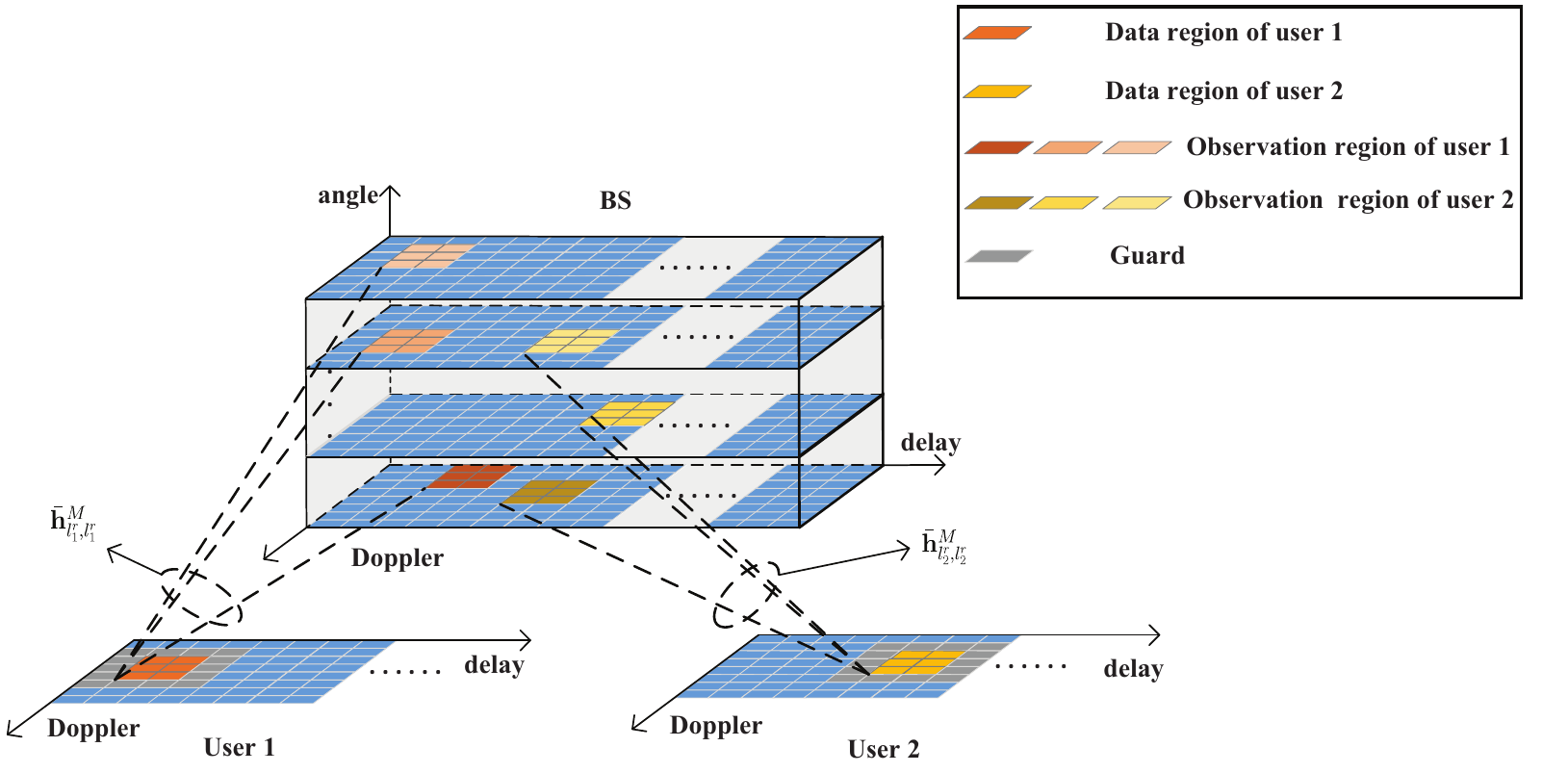}\\
  \caption{The channel dispersion over the angle-delay-Doppler domain for UL OTFS.}\label{fig_PDMA}
\end{figure*}

From \eqref{eq:fiveA Y^DD}, we can obtain that if only the $k$-th user sends only an effective symbol at $x_{k,i,j}$,
the BS will only receive the information of this symbol at $\bar{y}_{l, n, s}$ with the index set $(l= (i+i_{k,p})_{L_D}, n= \langle j+j_{k,p} \rangle, s = q_{k,p})$, where $(i_{k,p},j_{k,p},q_{k,p}) \in \mathcal Q_k$.
In other words, the OTFS scheme possesses the energy dispersion within the angle-delay-Doppler domain.
Once we achieve the scattering angle, the delays and the Doppler frequency, we can determine the exact dispersion locations within the OTFS block and collect the observed $\bar{y}_{l, n, s}$ at those grids to decode $x_{k,i,j}$.
Let us further consider two users and assume that the 3D channels for the $k_1$-th user and that for the $k_2$-th user are orthogonal in angle-delay-Doppler domain, which means that $\mathcal Q_{k_1} \cap \mathcal Q_{k_2} = \emptyset$.
Then, the $k_1$-th and $k_2$-th users can simultaneously send the data $x_{k_1,i,j}$ and $x_{k_2,i,j}$ at the same delay-Doppler grid.
BS can simultaneously extract different angle-delay-Doppler grids to recover $x_{k_1,i,j}$ and $x_{k_2,i,j}$ without inter-user interference.
However, if a number of data are densely placed in the delay-Doppler domain, many dispersed grids may overlap over the angle-delay-Doppler domain and cause severe inter-symbol and inter-user interference, which is the characteristic of OTFS.
Thus, the transmission requires powerful scheduling methods \cite{Ma_IA}.
Theoretically, we can schedule each symbol of users over the delay-Doppler domain to achieve several parallel and interference-free subchannels for the links from the users to BS.
However, the system spectrum efficiency would be very low.
Therefore, in the next subsection, we fully exploit the spatial super resolution of the massive antennas at BS and propose the PDMA scheme to implement the parallel data transmission with limited inter-symbol interference (ISI) over the angle-delay-Doppler domain.
A brief illustration for the UL part of our PDMA scheme is shown in Fig.~\ref{fig_PDMA}.

\subsection{Path Scheduling Algorithm}

Without loss of generality, we assume that the DOA $\theta_{k,p}$ for each scattering path of the same user is different, which means that each path of the same user possesses different angle signatures. In other words, different paths are associated with different distinguished domain grids.
Before proceeding, we define the double circular shift operation with respect to any $M\times N$ matrix $\mathbf X
=[\mathbf x_0,\mathbf x_{1},\ldots,\mathbf x_{N-1}]$ as
\begin{align}
{\mathbf X}^{(r,c)}=\text{Circ}_{c}\left\{\text{Circ}_{r}\{\mathbf x_0\},\text{Circ}_{r}\{\mathbf x_1\}, \ldots, \text{Circ}_{r}\{\mathbf x_{N-1}\}\right\},
\end{align}
where $\mathbf x_{n}$ is the $M\times 1$ vector and $\text{Circ}\{\cdot\}$ denotes the element-wise circular operation.
Explicitly, with respect to $\mathbf x_{n}$, the result of $\text{Circ}_r\{\cdot\}$ is one $M\times 1$ vector with the
$m$-th element as $[x_n]_{(m-r)_{M}}$. Correspondingly, for the operand $\{\mathbf x_0,\mathbf x_1,\ldots,\mathbf x_{N-1}\}$, $\text{Circ}_c\{\cdot\}$ obtains an $M\times N$ matrix, whose $n$-th column is $\mathbf x_{(n-c)_{N}}$.
Then, we first assume that only the $k$-th user sends its data $\mathbf X_k$ to BS along UL.
Correspondingly, the received information at the BS about $\mathbf X_k$, i.e., ${\bar y}_{l,n,s}$,
would be sparsely distributed at the grids of the 3D cube area
$\{(l,n,s)|l\in[0,L_D-1], n\in[0,N_D-1], s\in[0,N_r-1]\}$, which could be readily checked from (\ref{eq:fiveA Y^DD}).
Obviously, this observed 3D cube signal space contains $N_r$ layers along the angle direction, and
each layer has $L_DN_D$ delay-Doppler grids.

\begin{figure}[htbp]
  \centering
  \subfigure[]
  {\label{fig:block_compact}
  \begin{minipage}{70mm}
  \centering
  \includegraphics[width=70mm]{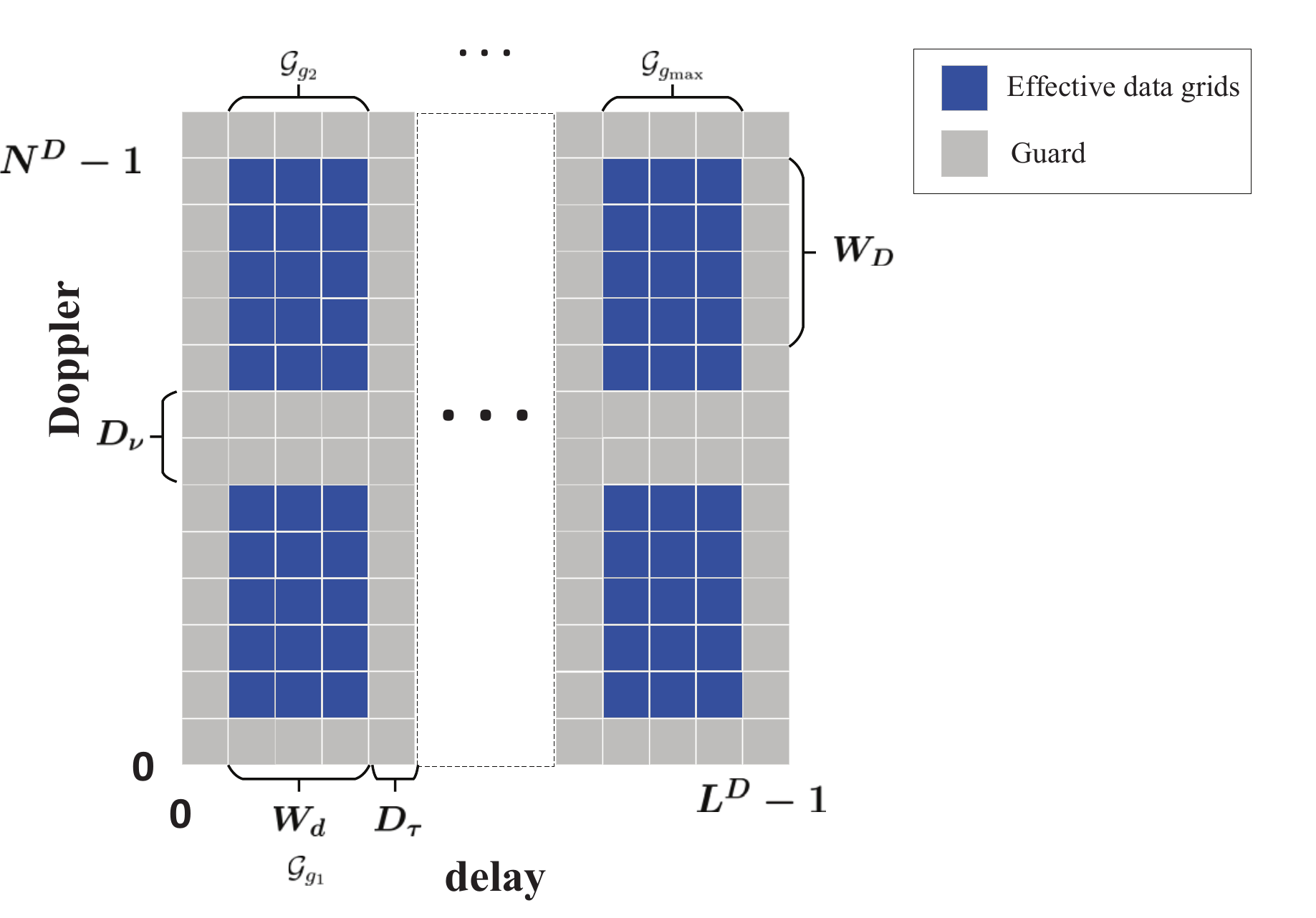}
  \end{minipage}
  }
  \subfigure[]
  {\label{fig:block_sparse}
  \begin{minipage}{70mm}
  \centering
  \includegraphics[width=70mm]{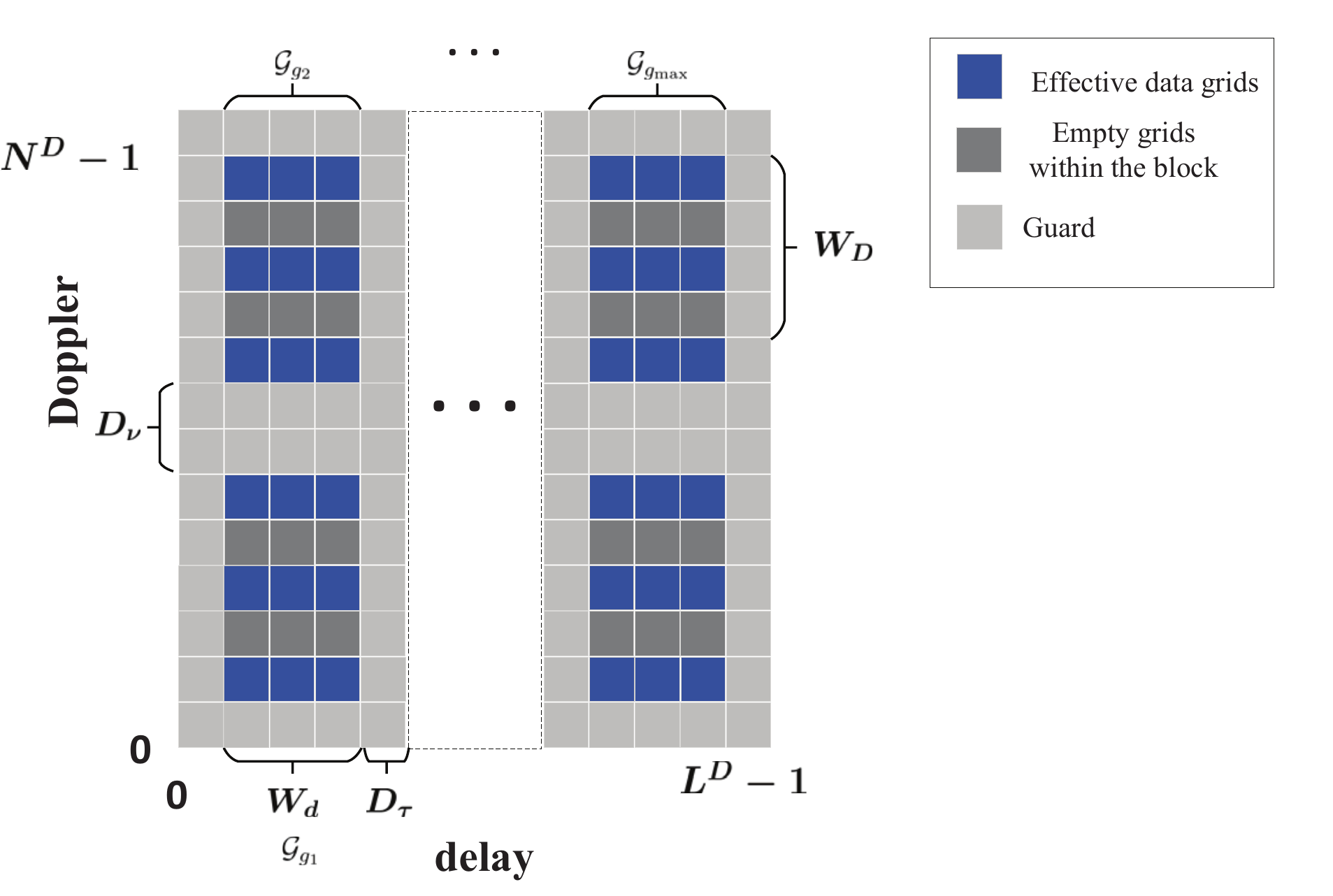}
  \end{minipage}
  }
  \caption{(a) Compact effective data block;
  (b) Sparse effective data block. }
\end{figure}

Let us define the $L_D\times N_D$ matrix $\overline {\mathbf Y}_s$ and $\overline{\mathbf W}_s$ whose $(l,n)$-th elements are $\bar y_{l,n,s}$ and $\bar w_{l,n,s}$, respectively.
Moreover, we define the diagonal matrices
$\mathbf {\overline H}_{k,(i,j,q)}={\bar{h}}_{k,(i, j, q)}\mathbf I_{L_D}$ and
$\mathbf{\overline G}_{k,(i,j,q)}^{(\ell)} = \text{diag}\left\{ \text{Circ}_{\ell} \{\mathbf g_{k,(i,j,q)} \}\right\}$,
where $\mathbf g_{k,(i,j,q)} = \left[{\bar{g}}_{k,(i, j, q)}^{0},{\bar{g}}_{k,(i, j, q)}^{1},\ldots,
{\bar{g}}_{k,(i, j, q)}^{L_D-1}\right]^T$.
With the results in the previous subsections, the energy of $\bar y_{l,n,s}$ are concentrated in $P$ matrices $\{\overline{\mathbf Y}_{q_{k,p}+N_r/2}\}_{p=1}^{P}$.
With (\ref{eq:fiveB H^DDA})--(\ref{eq:fiveA Y^DD}), we have
\begin{align}\label{receive_mtx}
\overline{\mathbf Y}_{q_{k,p}+N_r/2}
\!\!=&
\sum_{j'=0}^{N_D-1}
\left({\overline{\mathbf H}}_{k, (i_{k,p}, \langle j_{k,p}+j'\rangle, q_{k,p})} +
{\overline{\mathbf G}}_{k, (i_{k,p}, \langle j_{k,p}+j'\rangle, q_{k,p})}^{(0)} \right)
{\mathbf X}_k^{(i_{k,p}, j_{k,p}+j')}
+ {\overline{\mathbf W}}_{q_{k,p}+N_r/2},
\end{align}
where the above derivation utilizes the fact that there is no power leakage in the delay domain and all the paths of the $k$-th users can be separated by their angle information.
Obviously, the dominant signal on the right hand side  of \eqref{receive_mtx} is $\left({\overline{\mathbf H}}_{k, (i_{k,p},  j_{k,p}, q_{k,p})} + {\overline{\mathbf G}}_{k, (i_{k,p},  j_{k,p}, q_{k,p})}^{(0)} \right) {\mathbf X}_k^{(i_{k,p}, j_{k,p})}$, while the other parts can be the interference signal.
When all elements in $\mathbf X_k$ are effective data symbols with equal power,
the ratio between the dominant and the interference signals at $[\overline{\mathbf Y}_{q_{k,p}+N_r/2}]_{l,n}$
can be written as
\begin{align}
\Lambda_{l,n,{q_{k,p}+N_r/2}}
=&
\frac{\sum\limits_{n'=1}^{N_D-1} \sum\limits_{p'=1}^{P} \left|
\mathcal A(\boldsymbol \chi_{k,p'}, i_{k,p},\langle j_{k,p}+n'\rangle, q_{k,p}) \right|^2 \lambda_{k,p'}
}
{\sum\limits_{p'=1}^{P} \left|
\mathcal A(\boldsymbol \chi_{k,p'}, i_{k,p}, j_{k,p}, q_{k,p}) \right|^2 \lambda_{k,p'}
}\notag \\
\approx&
\sum\limits_{n'=1}^{N_D-1}
\sum\limits_{p'=1}^{P}
\frac{ \left|
\mathcal A(\boldsymbol \chi_{k,p'}, i_{k,p},\langle j_{k,p}+n'\rangle, q_{k,p}) \right|^2 \lambda_{k,p'}
}
{ \left|
\mathcal A(\boldsymbol \chi_{k,p'}, i_{k,p}, j_{k,p}, q_{k,p}) \right|^2 \lambda_{k,p'}
} \notag \\
=&
\sum\limits_{n'=1}^{N_D-1} \eta_{k,p}(n', 0),
\end{align}
where \eqref{eq:fiveB H^DDA} and \eqref{eq:fiveB g^DDA} are utilized, and $ \bar{ h}_{k,(p)} = \bar {h}_{k,(i_{k,p},j_{k,p},q_{k,p})}$ and $ \bar{ g}_{k,(p)} = \bar {g}_{k,(i_{k,p},j_{k,p},q_{k,p})}^{l}$ are defined for notational simplicity. With \eqref{interference_ratio}, it can be checked that $\eta_{k,p}(j, 0)$ monotonically decreases
with increasing $j$, and $\Lambda_{l,n,{q_{k,p}+N_r/2}}$ is mainly determined by $\eta_{k,p}(1, 0)$,
which is a small value.
Nonetheless, the data pattern in $\mathbf X_k$ can be further optimized to eliminate the power of
the interference signal. Simply, the non-zero effective data symbols can be equally placed along the Doppler direction.
Figs. \ref{fig:block_compact} and \ref{fig:block_sparse} illustrate the difference between the compact effective data blocks and the optimized data blocks.
Hence,  $\overline{\mathbf Y}_{q_{k,p}+N_r/2}$ can be approximated as
\begin{align}
\overline{\mathbf Y}_{q_{k,p}+N_r/2}
\!\!\approx&\left({\overline{\mathbf H}}_{k,(p)} + {\overline{\mathbf G}}_{k,(p)}^{(0)} \right) {\mathbf X}_k^{(i_{k,p}, j_{k,p})}
+ {\overline{\mathbf W}}_{q_{k,p}+N_r/2},
\end{align}
where  $\overline{\mathbf H}_{k,(p)}=\overline{\mathbf H}_{k,(i_{k,p},j_{k,p},q_{k,p})}$ and $\overline{\mathbf G}_{k,(p)}^{(0)} = \overline{\mathbf G}_{k,(i_{k,p},j_{k,p},q_{k,p})}^{(0)}$ are given for notational simplicity, which is because  $(i_{k,p},j_{k,p},q_{k,p})$ corresponds to the $p$-th path of the $k$-th user.

Without loss of generality, we restrict the effective data of user $k$ in one rectangle of $\mathbf X_k$, which
can be described by the grid set $\mathcal D_k$ as
\begin{align}\label{trans_mtx}
\mathcal D_{k} =& \left\{(l,n) | l \in [l_k, l_k+W_d-1], n \in [n_k, n_k+W_D-1] \right\},
\end{align}
where the element in $\mathcal D_{k}$ is the index of effective data. Moreover,
$l_k$ and $n_k$ separately denote the left and the bottom bounds of this rectangle, while
$W_d$ and $W_D$ are the maximum widths of the effective data region along delay and Doppler directions, respectively.
Then, for $(l,n)\in \mathcal D_k$, its received information at the BS would be distributed on $P$ layers of the 3D cubic area,
and the specific locations in $\mathbf Y_s$ can be written as $\{((l+i_{k,1})_{L_D}, \langle n+j_{k,1}\rangle, q_{k,1}+ N_r/2), ((l+i_{k,2})_{L_D}, \langle n+j_{k,2}\rangle, q_{k,2}+ N_r/2), \ldots, ((l+i_{k,P})_{L_D}, \langle n+j_{k,P} \rangle, q_{k,P}+ N_r/2)\}$.  Then, the received signal region with respect to $\mathcal D_k$ can be denoted as
\begin{align}\label{receive_set}
\mathcal C_{k} =& \left\{(l,n,s) | l \in [(l_k+i_{k,p})_{L_D}, (l_k+W_d-1+i_{k,p})_{L_D}], \right. \notag \\
&\left. n\in [\langle n_k+j_{k,p}\rangle, \langle n_k+W_D-1+j_{k,p}\rangle], s = q_{k,p}+N_r/2, p = 1,2,\ldots, P \right\}.
\end{align}
Correspondingly, we have
\begin{align}\label{rec_sig_small}
[{\overline{\mathbf Y}}_{q_{k,p}+N_r/2}^{(-i_{k,p}, -j_{k,p})}]_{\mathcal D_k}
=&
\left[({\overline{\mathbf H}}_{k,(p)} + {\overline{\mathbf G}}_{k,(p)}^{(-i_{k,p})} )\mathbf X_k \right]_{\mathcal D_k}
+ {\overline{\mathbf W}}_{q_{k,p}+N_r/2} \notag \\
&+ \sum_{j'=1}^{N_D-1} \left[({\overline{\mathbf H}}_{k,(i_{k,p}, \langle j_{k,p}+j'\rangle, q_{k,p})} + {\overline{\mathbf G}}_{k,(i_{k,p}, \langle j_{k,p}+j'\rangle, q_{k,p})}^{(-i_{k,p})} )\mathbf X_k \right]_{\mathcal D_k}.
\end{align}

Under the multi-user case, inter-user interference along UL is introduced.
Therefore, it is necessary to schedule $\mathcal D_k$ for each user and to {ensure that
the received effective data regions for different users do not overlap in the
3D cubic area, i.e.,
\begin{align}
\mathcal C_{k_1} \cap \mathcal C_{k_2}= \emptyset,
\end{align}
where $k_1\neq k_2$. To fully exploit the super resolution over the
angle domain, we schedule the users with two steps: within the first step, the users would be separated
over the angle domain, while in the second step, we further separate the users that have overlapped
angle signatures, in the delay-Doppler domain.
For the $k$-th user, its angle signatures are collected into the $\mathcal Q_k^a = \{q_{k,p}, p=1,2,\ldots, P\}$. Then, the users with non-overlapped angle signatures sets are allocated to the same group $\mathcal G_g$, i.e.,
\begin{align}
\mathcal Q_{k_1}^a \cap \mathcal Q_{k_2}^a = \emptyset,
\kern 10pt\text{and}~~
\text{dist}(\mathcal Q_{k_1}^a, \mathcal Q_{k_2}^a) \ge D_{\theta},
\end{align}
where $\text{dist}(\mathcal{Q}^a_{k_1}, \mathcal{Q}^a_{k_2}) \triangleq \min |q_{k_1,p}-q_{k_2,p}|, \forall q_{k_1,p} \in \mathcal{Q}^a_{k_1}, \forall q_{k_2,p} \in \mathcal{Q}^a_{k_2}$,
and the protection gap $D_{\theta}$ is added to mitigate the effect of the channel power leakage along
the angle direction. For $k_1, k_2\in \mathcal G_g$, we assign them the same delay-Doppler domain resource, i.e.,
$\mathcal D_{k_1}=\mathcal D_{k_2}=\mathcal D_{\mathcal G_g}$, but different angle grids, i.e., $\mathcal C_{k_1}\cap\mathcal C_{k_2}=\emptyset$, $k_1\neq k_2$.

With different user groups $\mathcal G_{g_1}$, $\mathcal G_{g_2}$, we assign distinguished delay-Doppler domain resources to
satisfy the following constraints:
\begin{align}
\mathcal D_{\mathcal G_{g_1}}\cap\mathcal D_{\mathcal G_{g_2}}=\emptyset
\kern 10pt\text{and}~~
\text{dist}(\mathcal D_{\mathcal G_{g_1}}, \mathcal D_{\mathcal G_{g_2}}) \succ \{D_\tau, D_\nu\}, \label{eq:user_group}
\end{align}
where $\text{dist}(\mathcal{D}_{\mathcal G_{g_1}}, \mathcal{D}_{\mathcal G_{g_2}})\succ \{D_\tau, D_\nu\} $ means either $\min |l_1-l_2|\geq D_\tau\ \text{or} \ \min |n_1-n_2|\geq D_\nu, \forall (l_1,n_1) \in \mathcal{D}_{\mathcal G_{g_1}}, \forall (l_2,n_2) \in \mathcal{D}_{\mathcal G_{g_2}}$,
and the guard gaps $D_{\tau}$ and $D_{\nu}$ are mainly utilized to combat the dispersion effect of
the 3D channels $\bar h_{k,(i,j,q)}$ along the delay and Doppler directions, respectively.
Define the maximum dispersion lengths along the delay and Doppler direction as
\begin{align}
D^{\max}_\tau = \max\{\{i_{k,p}\}_{p=1}^{P}\}_{k=1}^{K}, \kern 30pt
D^{\max}_\nu = \max\{\{|j_{k,p}|\}_{p=1}^{P}\}_{k=1}^{K}.\label{eq: user_group_delay}
\end{align}
Typically, we can separately set $D_{\tau}$ and $D_{\nu}$ as
$D_\tau=D^{\max}_\tau$ and $D_\nu=2D^{\max}_\nu$.
After scheduling, different users can
map their respective data to the scheduled delay-Doppler domain grids,
simultaneously send the data to the BS within
the same OTFS block, and occupy different 3D resources at the BS. Then, the BS can demap and decode different users' data in parallel without inter-user interference.

\subsection{3D Channel Reconstruction along UL}

In this subsection, we recover the equivalent 3D channel gains of the main and the secondary channels for data transmission.
According to (\ref{relation_h_g}), each user can send only one pilot
to estimate $\bar{h}_{k,(i,j,q)}$ and $\bar{g}_{k,(i,j,q)}^{\ell}$.
The position for the pilot of the $k$-th user in $\mathbf X_k$ is denoted as $(l_{k}^t,n_k^t)$,
$(l_{k}^t,n_k^t)\in\mathcal D_k$.
Then, its observed signal $[\mathbf {\overline Y}_{q_{k,p} + N_r/2}^{(-i_{k,p}, -j_{k,p})}]_{l_{k}^t,n_k^t}$ at the BS
can be written as
\begin{align}\label{receive_pilot_grid}
&[\mathbf {\overline Y}_{q_{k,p} + N_r/2}^{(-i_{k,p}, -j_{k,p})}]_{l_{k}^t,n_k^t}
=
e^{\jmath 2\pi \nu_{k,p} (l_k^{t}+i_{k,p})_{L_D} T_s }
{\bar{h}}_{k, (p)}
[\mathbf X_k]_{l_{k}^t,n_k^t}
+ [\mathbf{\overline W}_{q_{k,p} + N_r/2}]_{l_{k}^t,n_k^t} \notag \\
&\kern 6pt \!\!+\!\!
\sum_{(k',r')\neq (k,t)} \!\!
({\bar{h}}_{k', ((l_{k}^{t} \!+ i_{k,p} \!- l_{k'}^{r'})_{L_D}, \langle n_{k}^{t} \!+ j_{k,p} \!- n_{k'}^{r'} \rangle, q_{k,p})} \!\!+\!
{\bar{g}}_{k', ((l_{k}^{t} \!+ i_{k,p} \!- l_{k'}^{r'})_{L_D}, \langle n_{k}^{t} \!+ j_{k,p} \!-n_{k'}^{r'} \rangle, q_{k,p})}^{(l_k^t +i_{k,p})_{L_D}} ) [\mathbf X_{k'}]_{l_{k'}^{r'},n_{k'}^{r'}}
,
\end{align}
where \eqref{eq:fiveB H^DDA} and \eqref{eq:fiveB g^DDA} are utilized. With the LS estimator, we can recover ${\bar{h}}_{k, (p)}$ from the above equation as
\begin{align}\label{est_h_3D}
\widehat{\bar{h}}_{k,(p)}
= \frac{\Big[\mathbf {\overline Y}_{q_{k,p} + N_r/2}^{(-i_{k,p}, -j_{k,p})}\Big]_{l_{k}^t,n_k^t}}
{\Big[\mathbf X_k\Big]_{l_{k}^t,n_{k}^t} e^{\jmath 2\pi \nu_{k,p} (l_{k}^t+i_{k,p})_{L_D} T_s }}.
\end{align}
Correspondingly, the secondary channel associated with $\{\tau_{k,p}, \nu_{k,p}, \theta_{k,p}, h_{k,p}\}_{p=1}^P$ can be obtained from (\ref{relation_h_g}) as
\begin{align}
\widehat{\bar{g}}_{k,(p)}^{(l_{k}^t + i_{k,p})_{L_D}} \approx
2\jmath e^{\jmath \pi \nu_{k,p} (l_{k}^t+i_{k,p})_{L_D} T_s }
\sin(\pi \nu_{k,p} (l_{k}^t+i_{k,p})_{L_D} T_s)
\widehat{\bar{h}}_{k,(p)}.
\end{align}

However, as mentioned in Section II, the secondary channel $\bar g^\ell_{k,(i,j,q)}$ depends on
the specific position of the observed signal.
Furthermore, in order to enhance the data detection performance, we should
acquire the information about all the non-dominant 3D channels.
So, we give one 3D channel reconstruction scheme as follows.
With \eqref{eq:approx H^DDA} and the achieved $\{\tau_{k,p}, \nu_{k,p}, \theta_{k,p})\}_{p=1}^{P}$,
we can estimate $\{h_{k,p}\}_{p=1}^{P}$ as
\begin{align}\label{ori_h_est}
\hat{h}_{k,p} = \frac{{\sqrt N_r} \hat{\bar{h}}_{k,(p)} }
{ e^{\jmath2 \pi\nu_{k,p}T_s}
\mathcal A(\boldsymbol \chi_{k,p}, i_{k,p},j_{k,p},q_{k,p})}.
\end{align}

Until now, we can obtain the exact information about
$\{\tau_{k,p}, \nu_{k,p}, \theta_{k,p}, h_{k,p}\}_{p=1}^P$  within the current OTFS block.
Then, with \eqref{eq:fiveB H^DDA} and \eqref{eq:fiveB g^DDA}, we can reconstruct the 3D channels $\bar h_{k,(i,j,q)}$ and $\bar g_{k,(i,j,q)}^\ell$ at all grids in the 3D cubic area.


\subsection{UL Data Detection over the Angle-Delay-Doppler Domain}
Let us consider one specific data symbol in $[\mathbf X_k]_{\mathcal D_k}$, whose element index is denoted as $(l_k^r,n_k^r)\in \mathcal D_k$.
With \eqref{rec_sig_small}, it can be checked that the received information about
$[\mathbf X_k]_{l_k^r,n_k^r}$ at the BS lies in the 3D grid with the index in the set $\{((l_k^r+i_{k,p})_{L_D}, \langle n_k^r +j_{k,p} \rangle, q_{k,p}+ N_r/2)\}_{p=1}^{P}$.
Then, we collect the observations at the $P$ received grids of  $[\mathbf X_k]_{l_k^r,n_k^r}$ into a $P\times 1$
vector as
\begin{small}\begin{align}\label{receive_data_MMSE}
&\mathbf y_{l_k^r,n_k^r}
= \underbrace{\left[ \begin{matrix}
   {\bar{h}}_{k, (1)} +
{\bar{g}}_{k, (1)}^{(l_k^r+i_{k,1})_{L_D}} \\
   \vdots  \\
  {\bar{h}}_{k, (P)} +
{\bar{g}}_{k, (P)}^{(l_k^r+i_{k,P})_{L_D}}
\end{matrix} \right]}_{\mathbf h_{l_k^r,n_k^r}^{\text{M}}} \!
[\mathbf X_k]_{l_k^r,n_k^r}
+ \mathbf {\bar{w}}_{l_k^r,n_k^r} \notag \\
&\kern 12pt +\!\! \sum\limits_{(k',r')\neq(k,r)} \!\!
\underbrace{\left[ \begin{matrix}
   {\bar{h}}_{k', ((l_k^r \!+i_{k,1} \!-l_{k'}^{r'})_{L_D}, \langle n_{k}^{r} \!+j_{k,1} \!-n_{k'}^{r'} \rangle, q_{k,1})} \!\!+\!
{\bar{g}}_{k', ((l_k^r \!+i_{k,1} \!-l_{k'}^{r'})_{L_D}, \langle n_{k}^{r} \!+j_{k,1} \!-n_{k'}^{r'} \rangle, q_{k,1})}^{(l_{k}^{r}+i_{k,1})_{L_D}} \\
   \vdots  \\
   {\bar{h}}_{k', ((l_k^r \!+i_{k,P} \!-l_{k'}^{r'})_{L_D}, \langle n_{k}^{r} \!+j_{k,P} \!-n_{k'}^{r'} \rangle, q_{k,P})} \!\!+\!
{\bar{g}}_{k', ((l_k^r \!+i_{k,P} \!-l_{k'}^{r'})_{L_D}, \langle n_{k}^{r} \!+j_{k,P} \!-n_{k'}^{r'} \rangle, q_{k,P})}^{(l_{k}^{r}+i_{k,P})_{L_D}}
\end{matrix} \right]}_{\mathbf h_{{l_k^r,n_k^r},l_{k'}^{r'},n_{k'}^{r'}}^{\text{I}}}
[\mathbf X_{k'}]_{l_{k'}^{r'},n_{k'}^{r'}},
\end{align}\end{small}where $\mathbf {\bar{w}}_{l_k^r,n_k^r} = [[{\overline{\mathbf W}}_{q_{k,1}+N_r/2}]_{l_k^r,n_k^r}, \ldots, [{\overline{\mathbf W}}_{q_{k,P}+N_r/2}]_{l_k^r,n_k^r}]^T $ is the $P \times 1$ noise vector and the third term on the right hand of \eqref{receive_data_MMSE} is the sum of interferences.
Moreover, the $P\times 1$ vectors are defined in the above equation.
Hence, we make full use of the $P$ received signals and adopt
the MRC strategy for data detection.

From (\ref{receive_data_MMSE}), the signal-to-interference and noise ratio (SINR) along the $p$-th scattering path of the $k$-th
 user, i.e., ${\bar{h}}_{k, (p)} +
{\bar{g}}_{k, (1)}^{(l_k^r+i_{k,p})_{L_D}} $, can be expressed as
\begin{align}\label{SNR_pk}
\gamma_{k,p} =&
\frac{\left| [\mathbf h_{l_k^r,n_k^r}^{\text{M}}]_{p} \right|^2 }
{| \sum\limits_{(k',r)\neq (k,r')}
[\mathbf h_{{l_k^r,n_k^r},l_{k'}^{r'},n_{k'}^{r'}}^{\text{I}}]_{p}|^2
+ \sigma_n^2}
.
\end{align}

According to the MRC principle, the weighting factor for the signal along the $p$-th scattering path is $\gamma_{k,p}$ \cite{PDM}.
Define the SINR vector $\boldsymbol \gamma_k = [\gamma_{k,1}, \ldots, \gamma_{k,P}]^T$ and
combine the $P$ received signals with their own weighting factor,
then, the MRC received signal can be represented as
\begin{align}\label{data_MRC}
\bar{y}_{k,r}^{MRC} \!\!=& \boldsymbol \gamma_k^{\text {T}} \mathbf y_{l_k^r,n_k^r}
\notag \\
= & \!\!\sum_{p=1}^{P} \!\!\gamma_{k,p}
[\mathbf h_{l_k^r,n_k^r}^{\text{M}}]_{p}
[\mathbf X_k]_{l_{k}^{r},n_{k}^{r}}
\!\!+\!\! \sum_{p=1}^{P}\! \gamma_{k,p}\!\!
\sum_{(k',r')\neq (k,r)}\!
[\mathbf h_{{l_k^r,n_k^r},l_{k'}^{r'},n_{k'}^{r'}}^{\text{I}}]_{p}
[\mathbf X_{k'}]_{l_{k'}^{r'},n_{k'}^{r'}}
\!+\!\! \sum_{p=1}^{P}\! \gamma_{k,p} [\mathbf {\bar{w}}_{l_k^r,n_k^r}]_p,
\end{align}

From \eqref{data_MRC}, we can derive the optimized SINR of the combined signal as
\begin{align}\label{SINR_MRC}
\gamma_k^{MRC} \!\!=\!\! \frac{ \sum\limits_{p=1}^{P} \left| \gamma_{k,p}
[\mathbf h_{l_k^r,n_k^r}^{\text{M}}]_{p} \right|^2 }
{\sum\limits_{p=1}^{P}\sum\limits_{(k',r')\neq (k,r)} \left|
\gamma_{k,p}
[\mathbf h_{{l_k^r,n_k^r},l_{k'}^{r'},n_{k'}^{r'}}^{\text{I}}]_{p} \right|^2
+ \sum\limits_{p=1}^{P} \gamma_{k,p}^2 \sigma_n^2}.
\end{align}

As the optimal SINR is acquired, we can precisely recover the transmitted data with classical estimation algorithms, such as the LS detector. With the characteristic of LS, the mean square error (MSE) of the data detection is $\frac{1}{\gamma_k^{MRC}}$.

\subsection{DL Data Detection within the Delay-Doppler Domain}

In the DL, the BS's transmission resource over the angle-domain-Doppler domain can be described by $N_r$  matrices of size $L_D\times N_D$, i.e., $\mathbf X_s^d$, $s\in[0,N_r-1]$.
Then the BS can send data $[\mathbf X_{s}^d]_{l,n}$ within the 3D cubic area $\{(l,n,s)|l\in[0,L_D-1], n\in[0,N_D-1], s\in[0,N_r-1]\}$ and the users will receive their signal $[\mathbf Y_{k}^d]_{l,n}$ at the grids of the 2D area $\{(l,n)|l\in[0,L_D-1], n\in[0,N_D-1]\}$.
For the TDD system, due to the reciprocity between the UL/DL channels over the angle-delay-Doppler domain, the DL channel parameters $\{\tau_{k,p}, \nu_{k,p}, \theta_{k,p} \}_{p=1}^P$ are the same as the UL ones.
Moreover, the idea of the path scheduling algorithm along UL can be applied for the multi-user service along DL.
Then, with the angle grouping results in (\ref{eq:user_group}), the  received signal at the $k$-th user is mostly from the $P$ matrices, i.e.,  $\mathbf X_{q_{k,p}+ N_r/2}^d$, $p=1,2,\ldots, P$.
Correspondingly, the received signal $\mathbf Y_k^d$  can be obtained from \eqref{receive_mtx} as
\begin{align}\label{DL_all_mtx}
{\mathbf Y_k^d}
=& \sum_{p=1}^{P}
\sum_{j'=0}^{N_D-1}
\left({\overline{\mathbf H}}_{k, (i_{k,p}, \langle j_{k,p}+j'\rangle, q_{k,p})} +
{\overline{\mathbf G}}_{k, (i_{k,p}, \langle j_{k,p}+j'\rangle, q_{k,p})}^{(0)} \right)
\mathbf X_{q_{k,p}+N_r/2}^{d(i_{k,p}, j_{k,p}+j')}
+ {\mathbf N_k},
\end{align}
where $\mathbf N_k$ is the noise matrix whose elements are Gaussian distributed with zero mean and covariance $\sigma_n^2$.

Without loss of generality, the $k$-th user's effective observation region along DL can be set as its transmission region along UL, i.e., $\mathcal D_k$.
Then, with (\ref{DL_all_mtx}), the $k$-th user's transmitting data area over the 3D transmission resource space of the BS should be $\mathcal C_{k}^{dl} = \{((l-i_{k,p})_{L_D}, \langle n-j_{k,p}\rangle, q_{k,p}+N_r/2)|(l,n)\in\mathcal D_k, p\in[1,P]\}$, which is different from $\mathcal C_k$  for a given $\mathcal D_k$.
Nonetheless, it can be checked that the path scheduling results from (\ref{eq:user_group}) and (\ref{eq: user_group_delay}) can make sure that
\begin{align}
\mathcal C_{k_1}^{dl}\cap \mathcal C_{k_2}^{dl}=\emptyset,~k_1\neq k_2,
\end{align}
which means that the  transmission resources for different users are orthogonal over the angle-delay-Doppler domain.  Then,
the path scheduling results for UL can be directly applied for DL.
After path division,  the $k$-th user's grid $(l,n) \in \mathcal D_k$ observes the signal components from
$[\mathbf X_{q_{k,1}+N_r/2}^d]_{(l-i_{k,1})_{L_D}, \langle n-j_{k,1}\rangle}$,
$\ldots$, $[\mathbf X_{q_{k,P}+N_r/2}^d]_{(l-i_{k,P})_{L_D}, \langle n-j_{k,P}\rangle}$, which separately experience the 3D channels ${\bar{h}}_{k, (1)} +
{\bar{g}}_{k, (1)}^{l}$, ${\bar{h}}_{k, (2)} +
{\bar{g}}_{k, (2)}^{l}$, $\ldots$, ${\bar{h}}_{k, (P)} +
{\bar{g}}_{k, (P)}^{l}$.
Before proceeding, we give the following equation
\begin{align}
[\mathbf X_{q_{k,p}+N_r/2}^d]_{(l-i_{k,p})_{L_D}, \langle n-j_{k,p} \rangle} = [\mathbf b_{k,l,n}]_{p} u_{k,l,n},
\end{align}
where the $P \times 1$ vector $\mathbf b_{k,l,n}$ represents the beamforming operation with respect to the observation grid $(l,n)$ in $\mathbf Y_k^{d}$, and $u_{k,l,n}$ denotes the desired effective data for the $k$-th user.
Then, with (\ref{DL_all_mtx}), we can write $[{\mathbf Y_k^d}]_{l,n}$ as
\begin{align}\label{receive_data_vec_grid}
[{\mathbf Y_k^d}]_{l,n}
=& (\mathbf h_{l,n}^{\text{M}})^T
\mathbf b_{k,l,n}
u_{k,l,n}
+ [\mathbf N_k]_{l,n} \notag \\
&+\!\!
\sum\limits_{j'=1}^{N_D-1}\!\! \sum_{p=1}^{P}
({\bar{h}}_{k, (i_{k,p}, \langle j_{k,p}+j'\rangle, q_{k,p})}
\!\!+\! {\bar{g}}_{k, (i_{k,p}, \langle j_{k,p}+j'\rangle, q_{k,p})}^{l})
[\mathbf X_{q_{k,p}+\!N_r/2}^d]_{(l-i_{k,p})_{L_D}, \langle n-j_{k,p}-j'\rangle},
\end{align}
where $\mathbf b_{k,l,n}$ can be designed under the minimum-MSE beamforming framework as
\begin{align}\label{MMSE_beamforming}
\mathbf b_{k,l,n} =&
\frac{(\mathbf h_{l,n}^{\text{M}})^{*}}
{\|\mathbf h_{l,n}^{\text{M}}\|}
=[e^{j 2\pi \nu_{k,1} l T_s}{\bar{h}}_{k, (1)}, \ldots, e^{j 2\pi \nu_{k,P} l T_s}{\bar{h}}_{k, (P)}]^H.
\end{align}
Notice that the parameters $\{\nu_{k,p}\}_{p=1}^P$ are achieved from the UL channel parameter extraction, while $\{\bar h_{k,(p)}\}_{p=1}^P$ can be obtained from the nearest UL OTFS block.
Moreover, after beamforming, the equivalent channel $(\mathbf h_{l,n}^{\text{M}})^T \mathbf b_{k,l,n}$ at all the grids in $\mathcal D_k$ are the same.
Thus, a single grid in $\mathcal D_k$ is needed to implement the estimation of $(\mathbf h_{l,n}^{\text{M}})^T \mathbf b_{k,l,n}$.

In order to describe the relationship among different parts of the proposed scheme,
the overall block diagram of the proposed strategy is illustrated in \figurename{ \ref{fig:diagram}}.

\begin{figure}[!t]
  \centering
  \includegraphics[width=140mm]{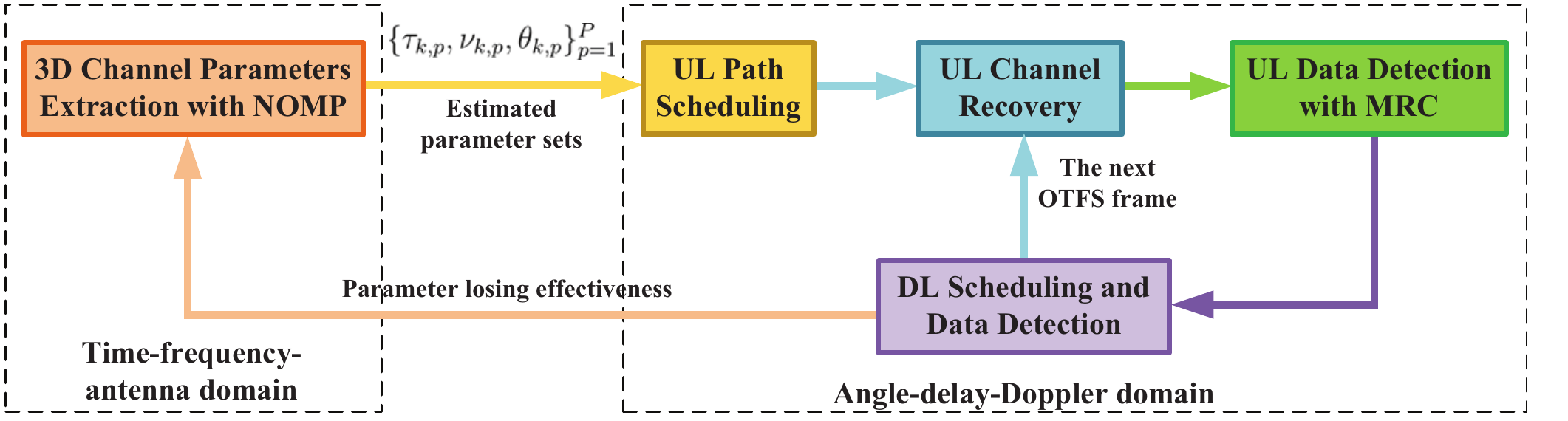}\\
  \caption{Diagram of the proposed scheme.} \label{fig:diagram}
\end{figure}

\vspace{-5mm}
\section{Simulation Results}
In this section, we evaluate the performance of our proposed algorithm for parameter capture and UL/DL data detection through numerical simulation.
Without loss of generality, we consider the TDD model.
$K=4$ is the number of users.
The number of BS's antennas is $N_r = 128$, the carrier frequency is 6 GHz, and the antenna spacing $d$ is set as half wavelength.
The users move at speed $v_s\in[120,360]$ km/h, and their maximal Doppler shift frequency is $2$ kHz.
With respect to the massive MIMO-OTFS scheme, the number of subcarriers is $L_D = 512$, the number of OFDM symbols in a single carrier within one OTFS block is $N_D = 128$, the length of CP is $L_{cp} = 32$, and the system sampling rate is $T_s = \frac{1}{20~\text{MHz}}$.
For the high mobility channel parameters, $\tau_{k,p}$ is randomly chosen from $\{0,T_s,2T_s,\dots,15T_s\}$, while $\theta_{k,p}$ and $\nu_{k,p}$ are uniformly distributed within $[-90^{\circ},90^{\circ}]$ and $[-2\ \text{kHz},2\ \text{kHz}]$, respectively.
Moreover, the exponentially decaying power delay profile $\sigma_{h_{k,p}}^2=\sigma_c^2e^{-\frac{\tau_{k,p}}{\tau_{k,\max}}}$
is adopted for $h_{k,p}$, where $\tau_{k,\max}=\max\{\tau_{k,p}\}_{p=1}^P$ and the constant $\sigma_c^2$ is chosen such that the average channel power is normalized to unity.
Furthermore, in the parameter capturing stage, the under-sampling rates in the 3D-coarse searching are set as $\eta_{\theta}=2$ and $\eta_{\nu}=2$, while the over-sampling rates in the 3D-precise searching are set as $\rho_{\theta}=4$ and $\rho_{\nu}=4$.
In addition, the width of the effective data block along the delay and Doppler domains are $W_d =50 $ and $W_D = 100$, respectively.
Further, the guard gaps along the delay and Doppler directions are set as $D_{\tau}=D_{\tau}^{\max} = 16$ and $D_{\nu} = 2D_{\nu}^{\max} = 14$, respectively.

The SNR is expressed as $\text{SNR}=10\log_{10}\sigma_{\mathbf{t}}^2/\sigma_n^2\text{dB}$. Here, we use the normalized MSE for the UL channel parameters $\boldsymbol{\theta}_k$, $\boldsymbol{\tau}_k$, $\boldsymbol{\nu}_k$ and $\boldsymbol{h}_k$ as the performance metric, which is defined as
\begin{align}
\text{MSE}_\mathbf{x}=\mathbb{E}\left\{
\frac{\|\hat{\mathbf{x}}-\mathbf{x}\|^2}
{\|\mathbf{x}\|^2}\right\},
\mathbf{x}=\boldsymbol{\theta}_k,\boldsymbol{\tau}_k,
\boldsymbol{\nu}_k, \boldsymbol{h}_k,
\end{align}
with $\hat{\mathbf{x}}$ as the estimate of $\mathbf{x}$ and
the $p$-th element of $\mathbf{x}$ consisting of  $\theta_{k,p}$, $\tau_{k,p}$, $\nu_{k,p}$ and $h_{k,p}$.


Fig. \ref{fig:pl_SNR} shows the MSE performance for $\boldsymbol{\theta}_k$, $\boldsymbol{\nu}_k$, and $\mathbf{h}_k$ at different SNRs, where two mobility conditions, i.e., the velocity $v_s$ of 240 km/h and 360 km/h are taken into consideration.
As shown in Fig. \ref{fig:pl_SNR}, the MSE decreases almost linearly as the SNR increases. Besides, the larger the speed $v_s$ is, the lower the MSE curves of $\boldsymbol{\theta}_k$, $\boldsymbol{\nu}_k$, and $\mathbf{h}_k$ are.
This is because the Doppler shift increases with the user speed, and the value of the Doppler phase bias vector $\mathbf{v}(\nu_{k,p})$ becomes larger, which leads to better performance of parameter capturing.
On the other hand, it can be seen from Fig. \ref{fig:pl_SNR} that the performance of our proposed 3D-NOMP algorithm at 360 km/h is only $1\sim 2$ dB lower than that at 240 km/h, which shows the effectiveness and robustness of our proposed 3D-NOMP method and also indicates that the high-mobility channel has a positive impact on the performance of parameter capturing.

\begin{figure}[!t]
\centering
\includegraphics[width=80mm]{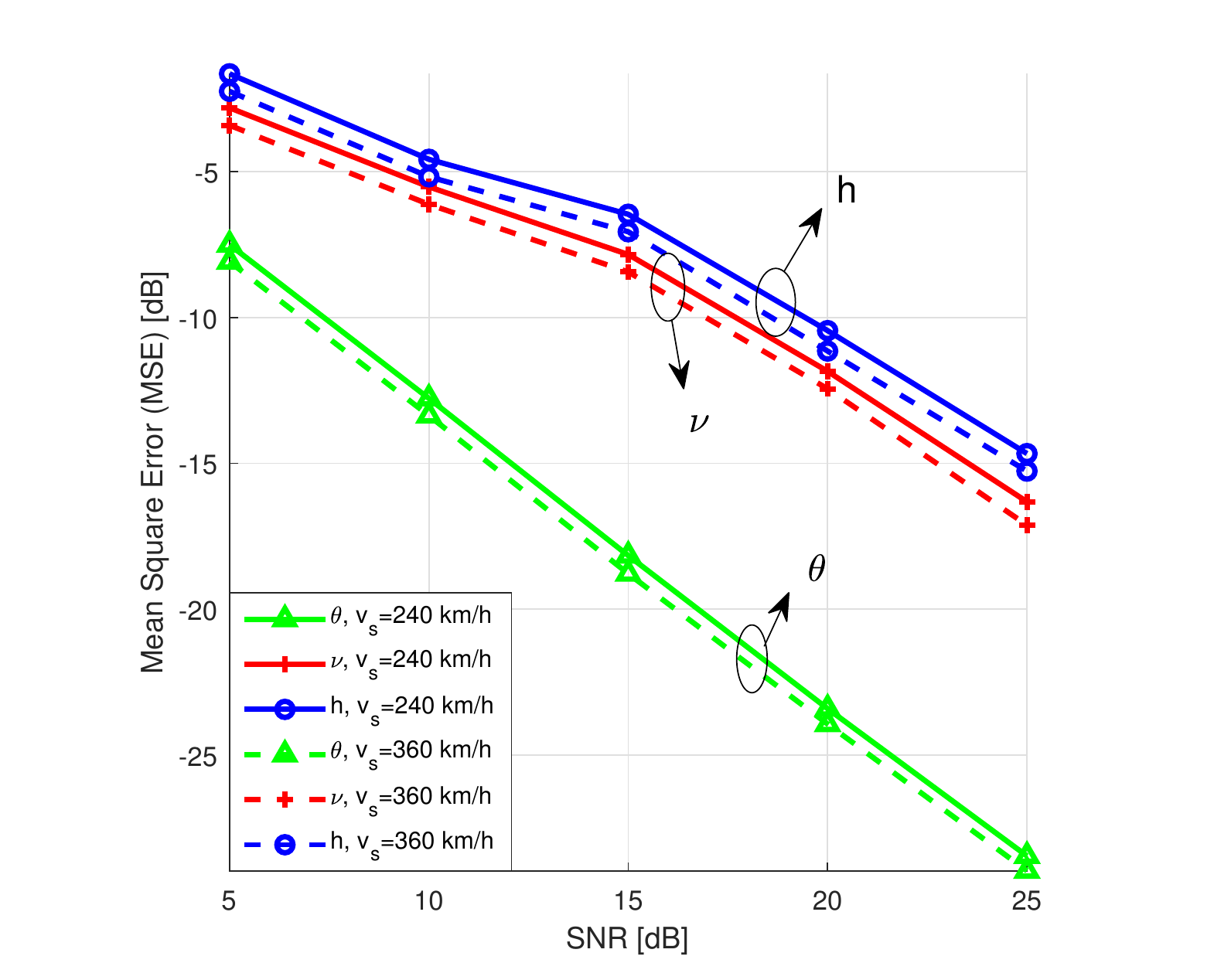}
\caption{The MSEs of $\boldsymbol{\theta}_k$, $\boldsymbol{\nu}_k$, and $\mathbf{h}_k$ versus SNR at different user's velocity versus SNR.}
\label{fig:pl_SNR}
\end{figure}

In Fig. \ref{fig:pl_N_t}, we study the MSE performance of channel parameters estimates with respect to the number of effective points $N_t$, where $\text{SNR}=20\ \text{dB}$.
It can be seen from Fig. \ref{fig:pl_N_t} that as $N_t$ increases, the MSEs for all parameters gradually decrease and converge at a tolerable value.
The above conclusion is not unexpected and can be explained as follows.
More observations help to distinguish the effect of Doppler shift on the channel, which in turn enhances the capture performance of other parameters.


Fig. \ref{fig:pl_New} shows the performance gain for the cyclic refinement of the extended Newton method with different number of iterations for the single refinement $R_s$.
Moreover, $N_t$ is fixed to 10 and SNR is set as 20 dB.
Without loss of generality, the iteration number of the cyclic refinement $R_c$ is set to 5.
From Fig.\ref{fig:pl_New}, it can be seen that both the single refinement and cyclic refinement scheme need only 4 iterations to converge, for all $\boldsymbol{\theta}_k$, $\boldsymbol{\nu}_k$ and $\mathbf{h}_k$.
Moreover, the MSEs of the estimated channel parameters for the single and cyclic refinement scheme is about 2 dB lower than that of the single refinement scheme, which indicates the  effectiveness of the proposed 3D-NOMP method.


\begin{figure}[htbp]
  \centering

  \begin{minipage}{65mm}
  \centering
  \includegraphics[width=65mm]{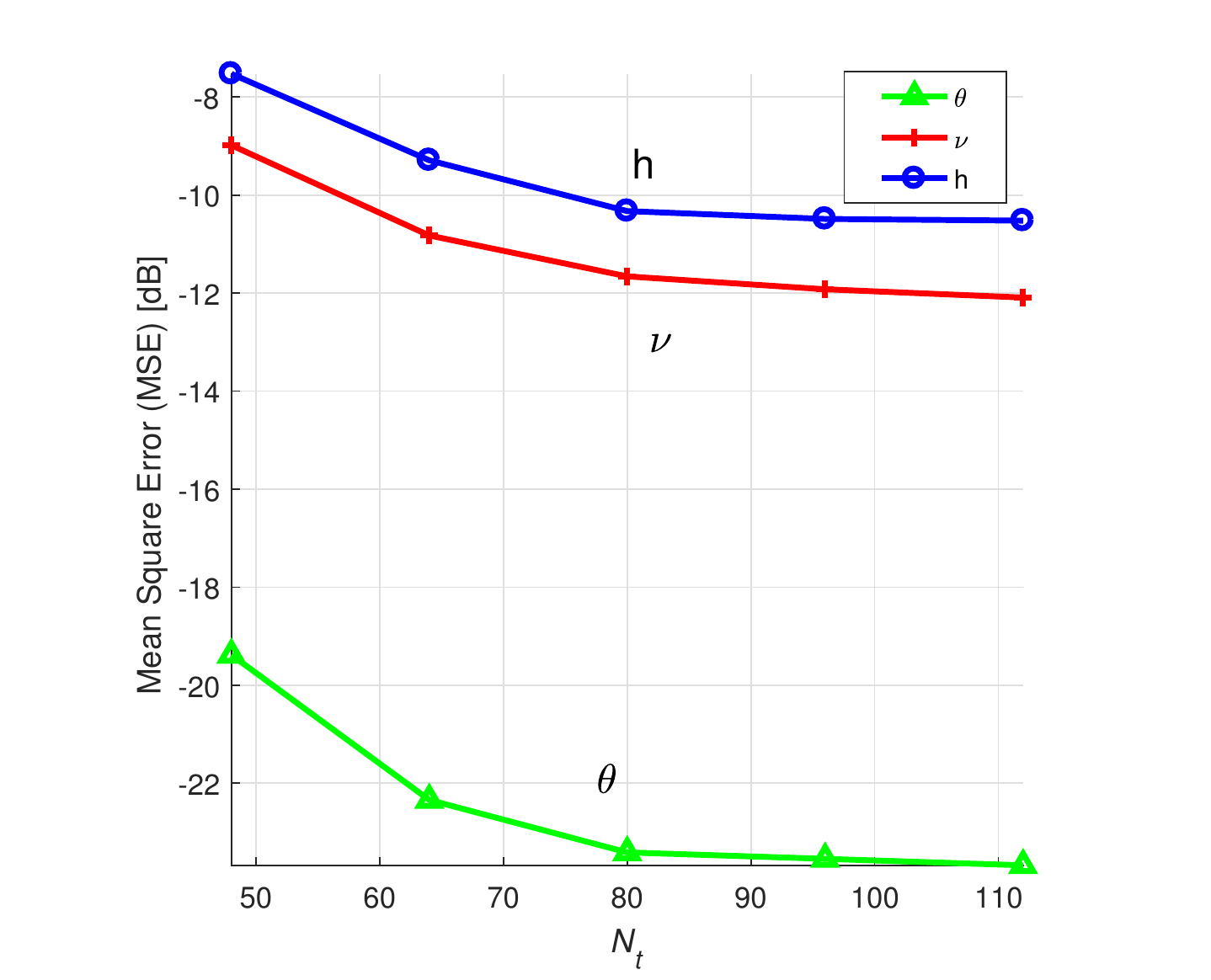}
  \caption{The MSEs of $\boldsymbol{\theta}_k$, $\boldsymbol{\nu}_k$, and $\mathbf{h}_k$ versus $N_t$.}
  \label{fig:pl_N_t}
  \end{minipage}
  \begin{minipage}{65mm}
  \centering
  \includegraphics[width=65mm]{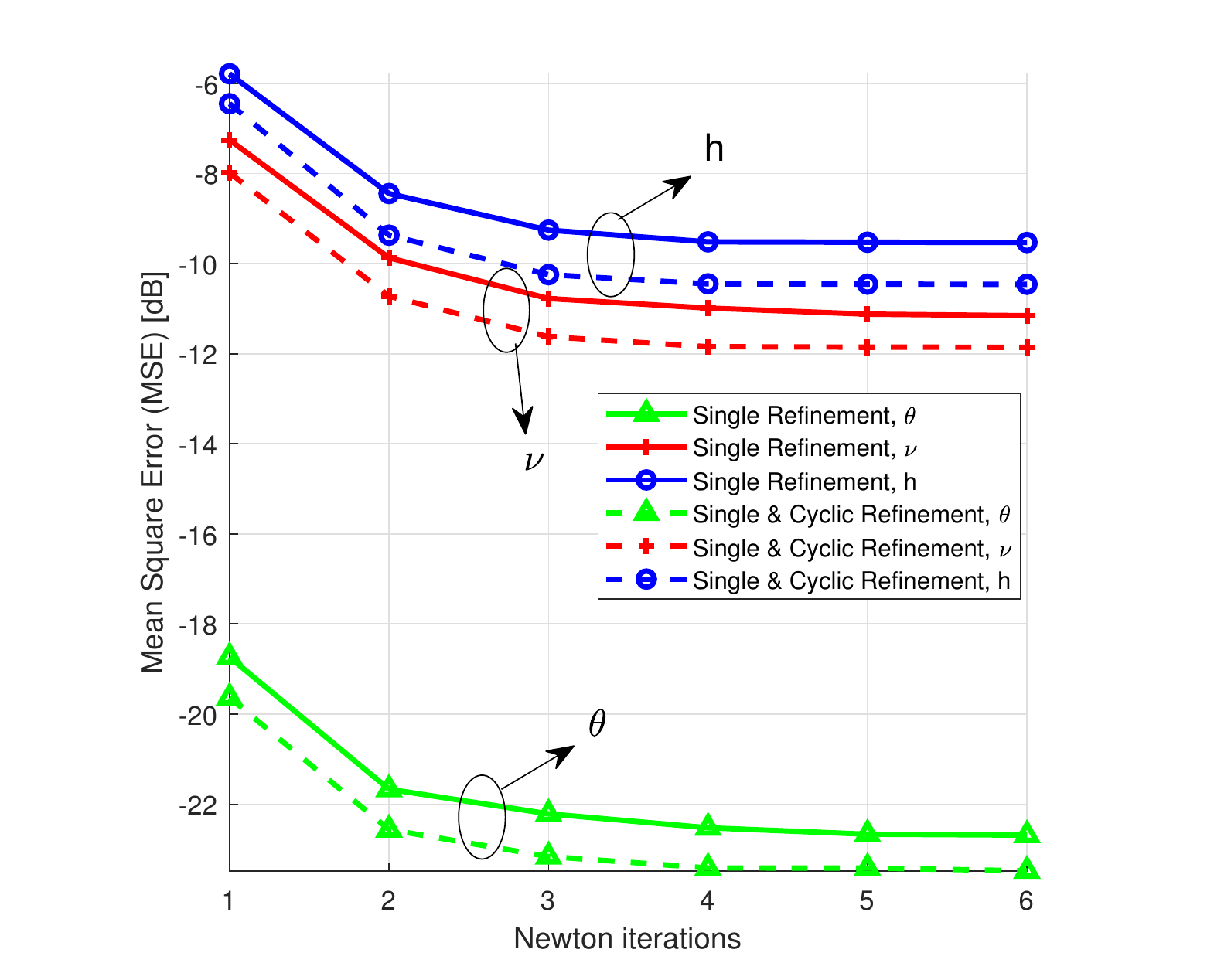}
  \caption{The MSEs of $\boldsymbol{\theta}_k$, $\boldsymbol{\nu}_k$, and $\mathbf{h}_k$ versus Newton iterations}.
  \label{fig:pl_New}
  \end{minipage}

\end{figure}


Figs. \ref{fig:ce_SNR_120} and \ref{fig:ce_SNR_360} show the comparison the channel recovery performance between lower and higher velocities.
The curves with triangles are the estimated 3D main channel without consideration of the secondary channel, while the curves with circles and that with crosses represent the estimated main channel and the secondary channel, respectively.
It can be seen that the MSE of the main channel without consideration of the secondary channel is much higher than the others.
Moreover, in the high mobility scenario, its performance worsens, while the other two curves are closer to those in lower mobility scenario.
This behaviour can be explained as the influence of the secondary channel increases by  increasing Doppler shift frequency.
In the high mobility case, the Doppler shift frequency becomes larger, and thus, the curve with triangles worsens.
For the other two curves, the estimated Doppler shift frequency in the 3D NOMP process is more accurate, which brings some gain for channel recovery.
Furthermore, the MSE of the secondary channel is always higher than that of the main channel, which is due to the approximation form of the estimating equation.

\begin{figure}[htbp]
  \centering
  \subfigure[]
  {\label{fig:ce_SNR_120}
  \begin{minipage}{70mm}
  \centering
  \includegraphics[width=70mm]{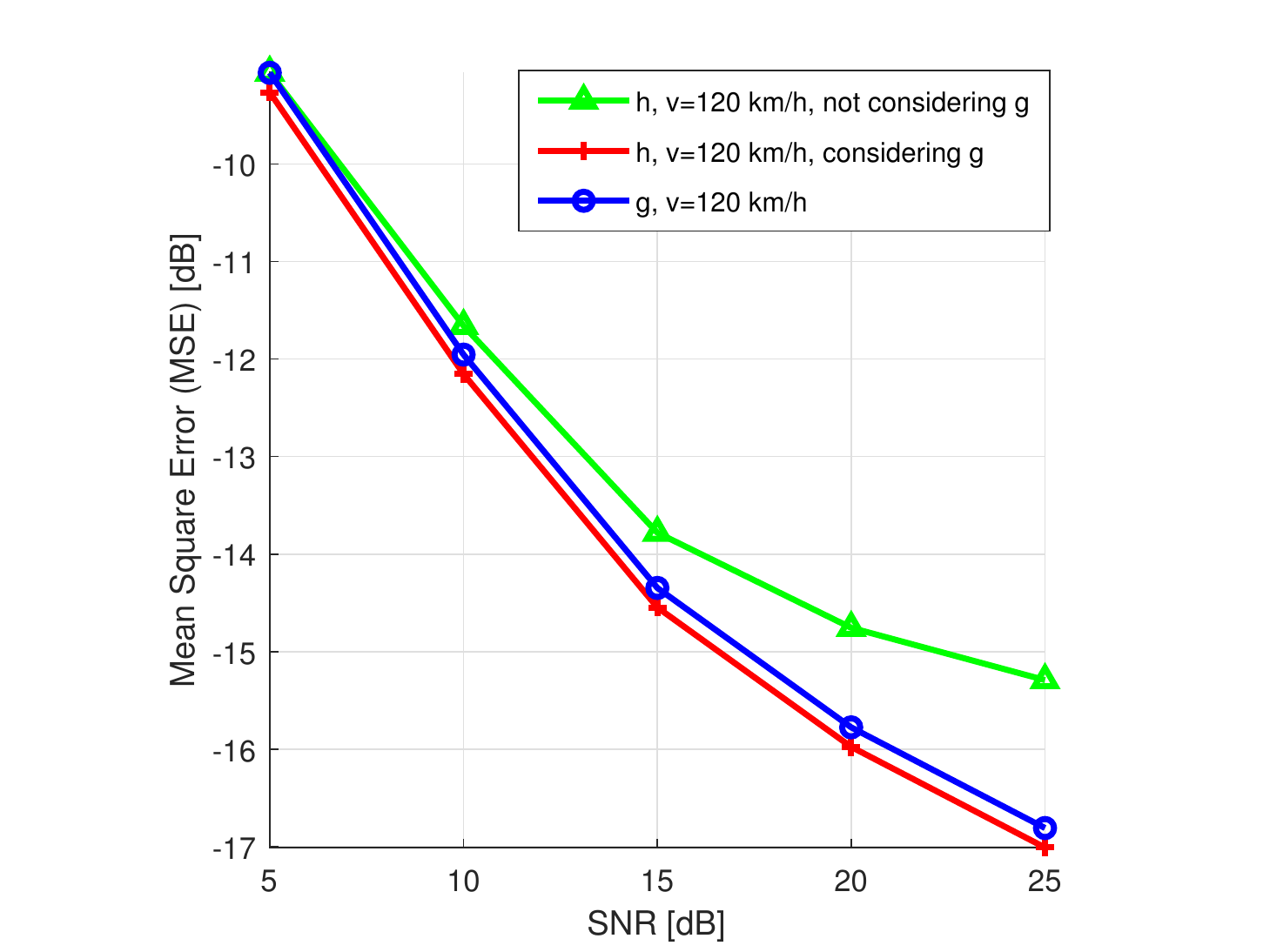}
  \end{minipage}
  }
  \subfigure[]
  {\label{fig:ce_SNR_360}
  \begin{minipage}{70mm}
  \centering
  \includegraphics[width=70mm]{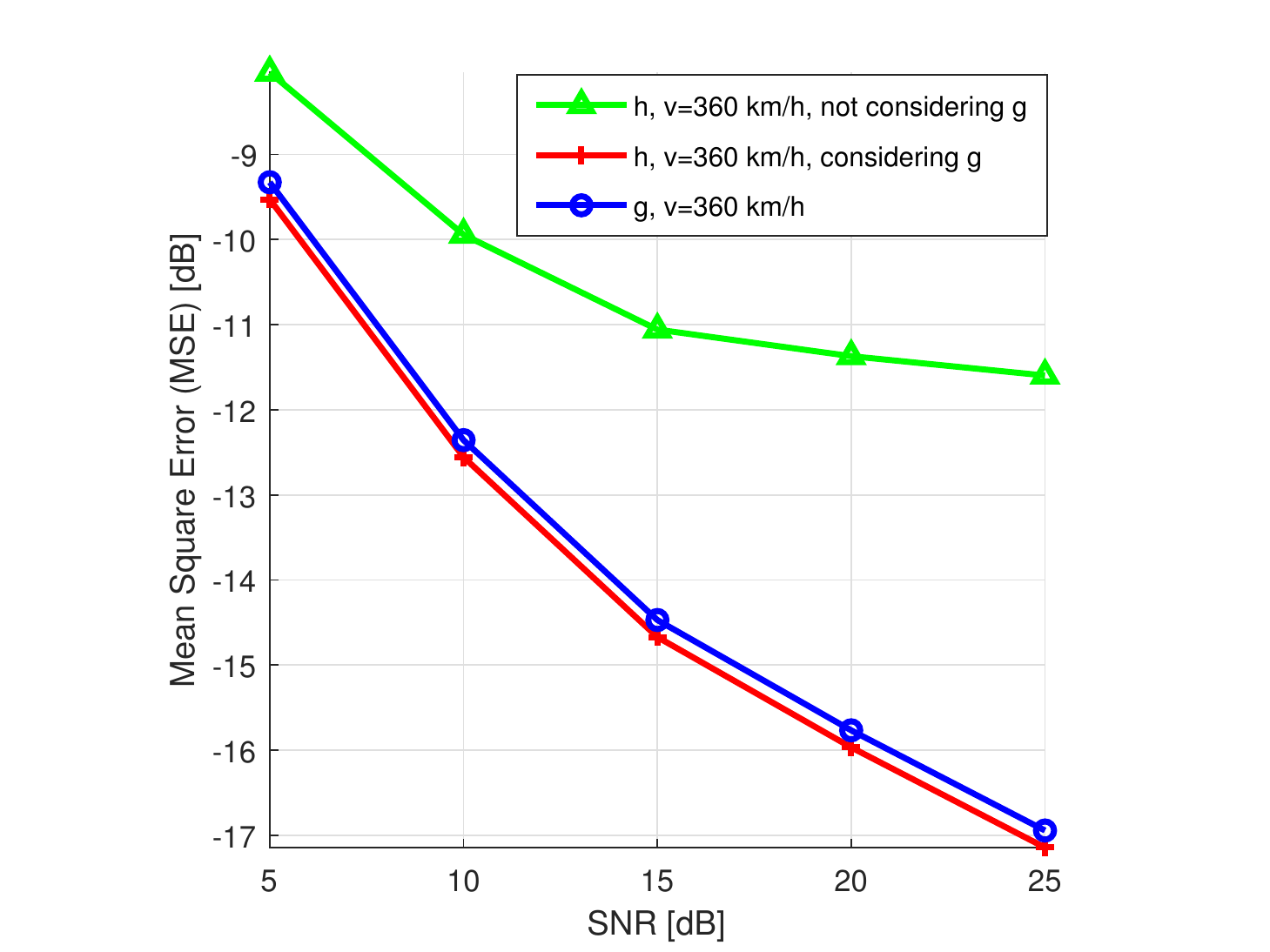}
  \end{minipage}
  }
  \caption{The MSEs of UL channel recovery with (a) $v_s = 120$ km/h;
  (b) $v_s = 360$ km/h. }
\end{figure}

Fig. \ref{fig:dd_SNR} depicts the data detection performance in two types of mobility cases, where the sparsity of the effective data block is considered.
As expected, the MSE decreases with increasing SNR, and tends to converge; the latter behaviour is due to the interference within the data block.
When the SNR is very low, the interference is very small compared with the noise.
However, at high SNR, the interference becomes much larger than the noise, and severely affects the data detection performance.
It is worth noting the gap between the MSE with sparsely scheduled data block and that with densely distributed data blocks.
This is due to the fact that the farther the dominant dispersed received grids are, the smaller the interference caused by the power leakage is.


Fig. \ref{fig:dl_SNR} illustrates the DL data detection performance versus SNR for different channel qualities and user velocities.
It can be observed that the MSE with perfect channel is much lower than that with mismatched channel gain, which is caused by the inaccurate estimated 3D channels.
Moreover, the MSE tends to converge at very high SNR, for the same reason mentioned for Fig.~\ref{fig:dd_SNR}.


\begin{figure}[htbp]
  \centering
  \begin{minipage}{70mm}
  \centering
  \includegraphics[width=70mm]{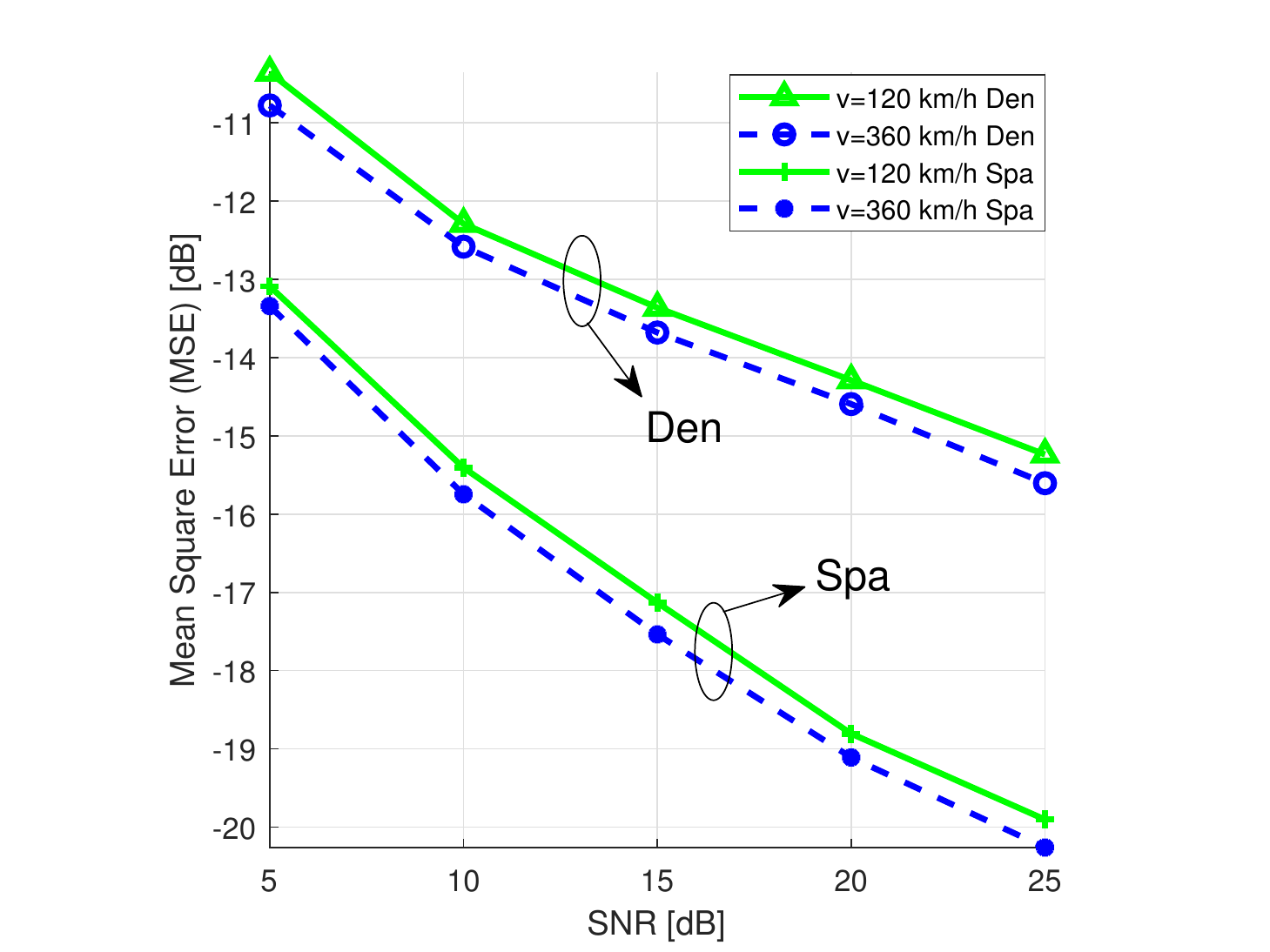}
  \caption{The MSEs of UL data detection versus SNR. Spa: sparsely scheduled data blocks; Den: densely distributed data blocks.}
  \label{fig:dd_SNR}
  \end{minipage}
  \begin{minipage}{70mm}
  \centering
  \includegraphics[width=70mm]{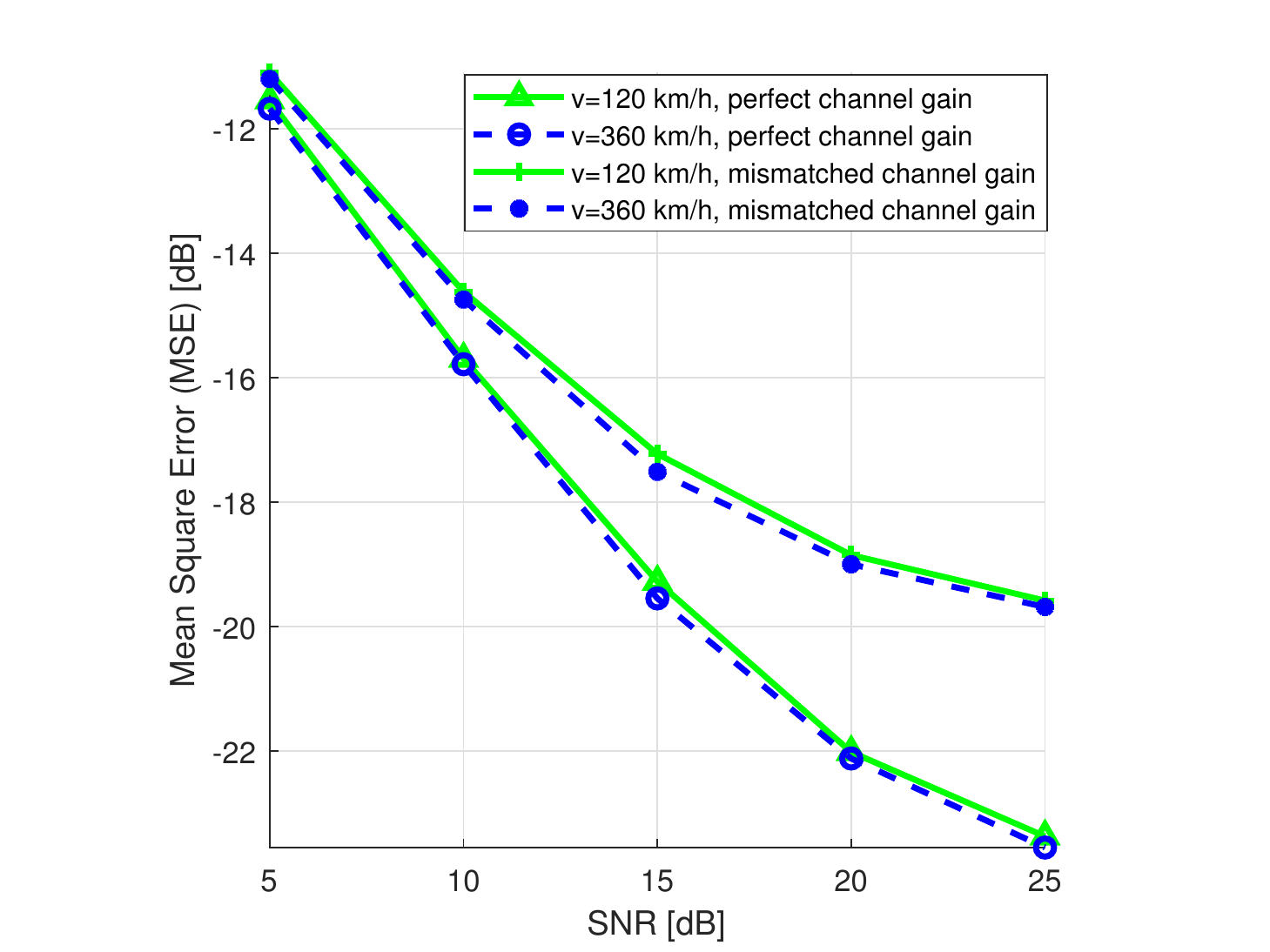}
  \caption{The MSEs of DL data detection versus SNR.}
  \label{fig:dl_SNR}
  \end{minipage}

\end{figure}

\vspace{-15pt}
\section{Conclusions}
In this paper, a new PDMA scheme for both the UL and DL massive MIMO-OTFS network was proposed.
The 3D UL channel model and the received signal model in the angle-delay-Doppler domain was first studied.
Then, the 3D-NOMP algorithm was utilized for the extraction of the 3D UL channel parameters.
After analyzing both energy dispersion and power leakage of the 3D channels,
a path scheduling algorithm was designed to properly assign the angle-domain resources at the user sides and to ensure the non-overlap of the observation regions for different users over the 3D cubic area.
By executing the scheduling algorithm, different users can independently map their respective data to the scheduled delay-Doppler domain grids, and simultaneously send the data to the BS without inter-user interference in the same OTFS block.
Moreover, the signals at desired grids within the 3D resource space of the BS are separately collected for the implementation of the 3D channel recovery and the MRC-based data detection.
Finally, a low-complexity beamforming scheme for the inter-user interference free DL communication was constructed.
Simulation results were provided to demonstrate the validity of our proposed unified UL/DL PDMA scheme.

\vspace{-4mm}
\section*{Appendix \\ Proof of Lemma 1}


\begin{proof}
From \eqref{eq:real Y^DD1}, using the time-varying channel model \eqref{eq:h_kln}, the received signal $[\mathbf Y_{n_r}]_{i, j+N_D/2} $ can be rewritten as
\begin{align}
[\mathbf Y_{n_r}]_{i, j+N_D/2}
= & \sum_{k=1}^{K} \sum_{i^{\prime}=0}^{L_D - 1} \sum_{j^{\prime} = -N_D / 2}^{N_D/2-1} [\mathbf X_k]_{i^{\prime}, j^{\prime}+N_D/2} \sum_{n=1}^{N_D}
\sum_{p=1}^{P} \frac{1}{N_D}h_{k, p} e^{\jmath 2 \pi \nu_{k, p} \left((n-1)\left(L_{D}+L_{c p}\right)+i+1\right) T_{s}} \notag \\
&\times
\delta\left((i-i^{\prime})_{L_D} T_{s}-\tau_{k, p}\right) e^{-\jmath2 \pi n_r \frac{d\sin \theta_{k,p}}{\lambda}}
e^{-j 2 \pi(n-1) \frac{j-j^{\prime}}{N_D}}
+ w_{i,j+N_D/2,n_r}. \label{ori_channel}
\end{align}

We can see from \eqref{ori_channel} that the original channel of each path is related to the position of the received grid, which is difficult for further path scheduling.
Moreover, by defining $\Upsilon_N (x) \triangleq \sum\limits_{n=1}^N e^{\jmath 2\pi \frac{x}{N}(n-1)} = \frac{\sin(\pi x)}{\sin(\pi \frac{x}{N})} e^{\jmath \pi \frac{x(N-1)}{N}} $, \eqref{ori_channel} can be further expressed as
\begin{align}
&[\mathbf Y_{n_r}]_{i, j+N_D/2} \notag \\
= & \sum_{k=1}^{K} \sum_{i^{\prime}=0}^{L_D - 1} \sum_{j^{\prime} = -N_D / 2}^{N_D/2-1} [\mathbf X_k]_{i^{\prime}, j^{\prime}+N_D/2}  \left(
\sum_{p=1}^{P} \frac{1}{N_D} h_{k, p} e^{\jmath 2 \pi \nu_{k, p} T_{s}}
\delta\left((i-i^{\prime})_{L_D} T_{s}-\tau_{k, p}\right) \right. \notag \\
&\left. \times \Upsilon_{N_D} (\nu_{k,p} N_D T\!-\!(j-j^{\prime}))
e^{-\jmath2 \pi n_r \frac{d\sin \theta_{k,p}}{\lambda}} + \sum_{p=1}^{P} (e^{\jmath 2\pi \nu_{k,p} i T_s }-1) \frac{1}{N_D} h_{k, p} e^{\jmath 2 \pi \nu_{k, p} T_{s}}
\right.\notag \\
&\left. \times \delta\left((i-i^{\prime})_{L_D} T_{s}-\tau_{k, p}\right)
\Upsilon_{N_D} (\nu_{k,p} N_D T\!-\!(j-j^{\prime}))
e^{-\jmath2 \pi n_r \frac{d\sin \theta_{k,p}}{\lambda}}   \right)
+ w_{i,j+N_D/2,n_r} \notag \\
= & \sum_{k=1}^{K} \!\! \sum_{i^{\prime}=0}^{L_D - 1} \!\! \sum_{j^{\prime} = -N_D / 2}^{N_D/2-1} \![\mathbf X_k]_{i^{\prime}, j^{\prime}\!+\!N_D/2} \!\times \! \left(\tilde{h}_{k,(i-i^{\prime})_{L_D},\langle j-j^{\prime}\rangle,n_r}
\!+\! \tilde{g}_{k,(i-i^{\prime})_{L_D},\langle j-j^{\prime}\rangle,n_r}^{i} \!\right)+ w_{i,j+N_D/2,n_r},
\end{align}
with $\tilde{h}_{k,i,j,n_r} $ and $\tilde{g}_{k,i,j,n_r}^{\ell} $ as defined in \eqref{eq:main H^DDS} and \eqref{eq:auxiliary H^DDS}, respectively.
\end{proof}

\linespread{1.16}

\balance


\begin{thebibliography}{1}
   \bibitem{massive_MIMO}
    T.~L.~Marzetta, ``Noncooperative cellular wireless with
    unlimited numbers of base station antennas,'' \emph{{IEEE} Trans. Wireless Commun.},
    vol.~9, no.~11, pp.~3590--3600, Nov. 2010.
	
	\bibitem{Massive_in_5G_1}
    V.~Jungnickel, K.~Manolakis, W.~Zirwas, B.~Panzner, V.~Braun, M.~Lossow, 	M.~Sternad, R.~Apelfrojd, and T.~Svensson, ``The role of small cells, coordinated multipoint, and massive {MIMO} in 5{G},'' \emph{{IEEE} Commun. Mag.}, vol.~52, no.~5, pp. 44--51, May 2014.

    \bibitem{4}
    S.~Jin, MR.~McKay, C.~Zhong, and K.~K.~Wong, ``Ergodic capacity analysis of amplify-and-forward MIMO dual-hop systems,'' \emph{IEEE Trans. Inf. Theory}, vol.~56, no.~5, pp.~2204--2224, May 2010.

    \bibitem{achie_rate}
    Y.~Xue, J.~Zhang, S.~Jin, and X.~Gao, ``On achievable rate of massive MIMO multiple access channels via virtual representation,'' \emph{Physical Communication}, vol.~20, pp.~133--140, Sep. 2016.

    \bibitem{2}
    L.~Fan, S.~Jin, C.~K.~Wen, and H.~Zhang, ``Uplink achievable rate for massive MIMO systems with low-resolution ADC,''  \emph {IEEE Communications Lett.}, vol. 19, no. 12, pp. 2186--2189, Dec. 2015.

    \bibitem{interference_quanti}
    J.~Chen and V.~K.~N.~Lau, ``Two-tier precoding for FDD multi-cell
    massive MIMO time-varying interference networks,''  \emph{{IEEE} J. Sel.
    Areas Commun.}, vol.~32, no.~6, pp.~1230--1238, Jun. 2014.


    \bibitem{energy_efficiency}
    W.~Tan, M.~Matthaiou, S.~Jin, and X.~Li, ``Spectral efficiency of DFT-based
    processing hybrid architectures in massive MIMO,'' \emph{{IEEE} Wireless
    Commun. Lett.}, vol.~6, no.~5, pp.~586--589, Oct. 2017.

    \bibitem{beamforming}
    J.~Mao, Z.~Gao, Y.~Wu, and M.~Alouini, ``Over-sampling codebook-based hybrid minimum sum-mean-square-error precoding for millimeter-wave 3D-MIMO,'' \emph{{IEEE} Wireless Commun. Lett.}, vol.~7, no.~6, pp.~938--941, Dec. 2018.

    \bibitem{user_grouping}
    A.~Adhikary, E.~AI Safadi, M.~K.~Samimi, R.~Wang, G.~Caire, T.~S.~Rappaport and A.~F.~Molisch, ``Joint spatial division and multiplexing for mm-Wave channels,''  \emph{{IEEE} J. Sel. Areas Commun.}, vol.~32, no.~6, pp.~1239--1255, Jun. 2014.

    \bibitem{data_detection}
    J.~Chen, ``A low complexity data detection algorithm for uplink multiuser massive MIMO systems,''  \emph{{IEEE} J. Sel. Areas Commun.}, vol.~35, no.~8 pp.~1701--1714, Aug. 2017.

    \bibitem{angle_domain}
    D.~Fan, F.~Gao, G.~Wang, Z.~Zhong, and A.~Nallanathan, ``Angle domain signal processing-aided channel estimation for indoor 60-GHz TDD/FDD massive MIMO systems,''  \emph{{IEEE} J. Sel. Areas Commun.}, vol.~35, no.~9, pp.~1948--1961, Sep. 2017.

    \bibitem{channel_recovery2}
    L.~You, X.~Gao, A.~L.~Swindlehurst, and W. Zhong, ``Channel acquisition for massive MIMO-OFDM with adjustable phase shift pilots.'' \emph{{IEEE} Trans. Signal Process.}, vol.~64, no.~6, pp. 1461--1476, Mar. 2016.

    \bibitem{channel_recovery3}
    Z.~Gao, L.~Dai, Z.~Wang, and S.~Chen, ``Spatially common sparsity based adaptive channel estimation and feedback for FDD massive MIMO,'' \emph{{IEEE} Trans. Signal Process.}, vol.~63, no.~23, pp.~6169--6183, Dec. 2015.

    \bibitem{channel_recovery_GaoZhen}
    A.~Liao, Z.~Gao, H.~Wang, S.~Chen, M.~Alouini, and H.~Yin, ``Closed-loop sparse channel estimation for wideband millimeter-wave full-dimensional MIMO systems,'' \emph{{IEEE} Trans. Commun.}, vol.~67, no.~12, pp.~8329--8345, Dec. 2019.

    \bibitem{gao_E}
	H.~Xie, F.~Gao, S.~Zhang, and S.~Jin, ``{A unified transmission strategy for
		{TDD/FDD} massive {MIMO} systems with spatial basis expansion model},''
	\emph{{IEEE} Trans. Veh. Technol.}, vol.~66, no.~4, pp. 3170--3184, Apr. 2017.

    \bibitem{Ma_J_SBL_Time_varing}
    J.~Ma, S.~Zhang, H.~Li, F.~Gao, and S.~Jin, ``{Sparse Bayesian learning for the time-varying massive MIMO channels: Acquisition and tracking},''
	\emph{{IEEE} Trans. Commun.}, vol.~67, no.~3, pp. 1925--1938, Mar. 2019.

    \bibitem{MYLi_SBL_Time_varing}
    M.~Li, S.~Zhang, N.~Zhao, W.~Zhang, and X.~Wang, ``{Time-varying massive MIMO channel estimation: Capturing, reconstruction and restoration},''
	\emph{{IEEE} Trans. Commun.}, vol.~67, no.~11, pp. 7558--7572, Nov. 2019.

    \bibitem{channel_recovery1}
    J.~Zhao, F.~Gao, W.~Jia, J.~Zhao, and W.~Zhang, ``Channel tracking for massive MIMO systems with spatial-temporal basis expansion model,'' in \emph{Proc. 2017 IEEE International Conference on Communications (ICC)}, Paris, May. 2017, pp.~1--5.

    \bibitem{Q_Qin}
    Q.~Qin, L.~Gui, B.~Gong, and S.~Luo, ``{Sparse channel estimation for massive MIMO-OFDM systems over time-varying channels},''
	\emph{{IEEE} Access}, vol.~6, pp.~33740--33751, Jul. 2018.

    \bibitem{You_OFDM}
    L.~You, X.~Gao, A.~L.~Swindlehurst, and W.~Zhong, ``{Channel acquisition for massive MIMO-OFDM with adjustable phase shift pilots},''
	\emph{{IEEE} Trans. Signal Process.}, vol.~64, no.~6, pp. 1464--1476, Mar. 2016.

    \bibitem{Octavia_OFDM}
    Y.~Zhang, R.~Venkatesan, O.~A.~Dobre, and C.~Li, ``Novel compressed sensing-based channel estimation algorithm and near-optimal pilot placement scheme,''
    \emph{{IEEE} Trans. Wireless Commun.}, vol.~15, no.~4, pp.~2590--2603, Apr. 2016.

    \bibitem{Timevarying_Guvensen}
    G.~M.~Guvensen and E.~Ayanoglu, ``Beamspace aware adaptive channel estimation for single-carrier time-varying massive MIMO channels,'' in \emph{Proc. 2017 IEEE International Conference on Communications (ICC)}, Paris, May. 2017, pp.~1--7.

    \bibitem{high_mobility4}
    Z.~Hu and W.~Zhang, ``Radiation efficiency aware high-mobility massive MIMO with antenna selection,'' \emph{{IEEE} Trans. Veh. Technol.}, vol.~68, no. 11, pp. 11363--11367, Nov. 2019.

    \bibitem{BEM1}
    J.~Zhao, F.~Gao, W.~Jia, S.~Zhang, S.~Jin, and H.~Lin, ``Angle domain hybrid precoding and channel tracking for millimeter wave massive MIMO systems,'' \emph{{IEEE} Trans. Wireless Commun.}, vol.~16, no.~10, pp.~6868--6880, Oct. 2017.

    \bibitem{BEM2}
    S.~Zhang, F.~Gao, J.~Li, and H.~Li, ``Time varying channel estimation for DSTC-based relay networks: Tracking, smoothing and BCRBs,'' \emph{{IEEE} Trans. Wireless Commun.}, vol.~14, no.~9, pp.~5022--5037, Sep. 2015.

    \bibitem{OTFS_original}
    R.~Hadani, S.~Rakib, S.~Kons, M.~Tsatsanis, A.~Monk, C.~Ibars, J.~Delfeld, A.~Goldsmith, A.~F.~Molisch, and R.~Calderbank, ``{Orthogonal time frequency space  modulation,}'' arXiv: 1808.00519v1, 2018. [Online]. Available: https://arxiv.org/abs/1808.00519.

   \bibitem{IC_OTFS}
    P.~Raviteja, K.~T.~Phan Y.~Hong, and E.~Viterbo, ``Interference cancellation and iterative detection for orthogonal time frequency space modulation,''
    \emph{{IEEE} Trans. Wireless Commun.}, vol.~17, no.~10, pp.~6501--6515, Oct. 2018.

    \bibitem{OTFS5}
    P.~Raviteja, K.~T.~Phan, and Y.~Hong, ``{Embedded pilot-aided channel estimation for OTFS in eelay-Doppler channels},'' arXiv:1808.08360v1, 2018. [Online]. Available: https://arxiv.org/abs/1808.08360.

    \bibitem{OTFS_MA}
    V.~Khammammetti and S.~K.~Mohammed, ``{OTFS-based multiple-access in high Doppler and delay spread wireless channels},'' \emph{{IEEE} Wireless Commun. Lett.}, vol.~8, no.~2, pp.~528--531, Apr. 2019.

    \bibitem{OTFS_dd1}
    W.~Yuan, Z.~Wei, J.~Yuan, and D.~W.~K.~Ng, ``{A simple variational Bayes detector for orthogonal time frequency space (OTFS) modulation},'' arXiv:1911.12538v1, 2019. [Online]. Available: https://arxiv.org/abs/1911.12538.

    \bibitem{OTFS_dd2}
    S.~Tiwari, S.~S.~Das, and V.~Rangamgari, ``{Low complexity LMMSE receiver for OTFS},'' arXiv:1910.01350v1, 2019. [Online]. Available: https://arxiv.org/abs/1910.01350.

    \bibitem{proof}
    W.~Shen, L.~Dai, J.~An, P.~Fan, and R.~W.~Heath, ``{Channel estimation for orthogonal time frequency space (OTFS) massive MIMO},'' \emph{{IEEE} Trans. Signal Process.}, vol.~67, no.~16, pp. 4204--4217, Aug. 2019.

    \bibitem{OTFS_model}
    G.~D.~Surabhi, R.~M.~Augustine, and A.~Chockalingam, ``{Peak-to-average power ratio of OTFS modulation},'' \emph{{IEEE} Commun. Lett.}, vol.~23, mo.~6, pp.~999--1002, Jun. 2019.

    \bibitem{ori_NOMP}
    B.~Mamandipoor, D.~Ramasamy, and U.~Madhow, ``{Newtonized orthogonal matching pursuit: Frequency estimation over the continuum},''	\emph{{IEEE} Trans. Signal Process.}, vol.~64, no.~19, pp.~5066--5081, Oct. 2016.

    \bibitem{Jin_NOMP}
    Y.~Han, T.~Hsu, C.~Wen, K.~Wong, and S.~Jin, ``{Efficient downlink channel reconstruction for FDD multi-antenna systems},'' \emph{{IEEE} Trans. Signal Process.}, vol.~18, no.~6, pp.~3161--3176, Jun. 2019.

    \bibitem{Ma_IA}
    J.~Ma, S.~Zhang, H.~Li, N.~Zhao, and V.~C.~M.~Leung, ``{Interference-alignment and soft-space-reuse based cooperative transmission for multi-cell massive MIMO networks},'' \emph{{IEEE} Trans. Wireless Commun.}, vol.~17, no.~3, pp.~1907--1922, Mar. 2018.

    \bibitem{PDM}
    Y.~Zeng and R.~Zhang, ``{Millimeter wave MIMO with lens antenna array: A new path division multiplexing paradigm},'' \emph{{IEEE} Trans. Commun.}, vol.~64, no.~4, pp. 1557--1571, Apr. 2016.

\end{thebibliography}
\end{document}